\RequirePackage{lineno}
\documentclass[twocolumn,showpacs,superscriptaddress,amsmath,amssymb,nofootinbib]{revtex4-2}

\usepackage{graphicx}
\usepackage{subfigure}
\usepackage{subcaption}
\usepackage{epsfig}
\usepackage{overpic}
\usepackage{dcolumn}
\usepackage{ulem}
\usepackage{bm}
\usepackage{color}
\usepackage{lineno}
\usepackage{xspace}
\usepackage{multirow}
\usepackage{epstopdf}
\usepackage{xcolor}
\usepackage{soul}
\usepackage{verbatim}
\usepackage{enumitem}
\usepackage{todonotes}
\usepackage{cancel}
\usepackage{array}
\usepackage{balance}
\usepackage[toc,page]{appendix}
\usepackage[colorlinks,allcolors=black]{hyperref}
\usepackage{array}
\usepackage{tabularray}
\UseTblrLibrary{booktabs, siunitx} 
\usepackage{tabularx}
\usepackage{footmisc}
\DefineFNsymbols*{myfns}{{*}{†}{‡}{§}{¶}}
\setfnsymbol{myfns}

\usepackage{diagbox}

\newcommand{\moe}{\affiliation{Key Laboratory of Atomic and Subatomic Structure and Quantum Control (MOE), Guangdong-Hong Kong Joint Laboratory of Quantum Matter, Guangzhou 510006, China
}}

\newcommand{\sfim}{\affiliation{Guangdong Basic Research Center of Excellence for Structure and Fundamental Interactions of Matter, Guangdong Provincial Key Laboratory of Nuclear Science, Guangzhou 510006, China}}

\newcommand{\ihep}{\affiliation{Institute of High Energy Physics, Chinese Academy of Sciences, Beijing 100049, China}}

\newcommand{\iqm}{\affiliation{State Key Laboratory of Nuclear Physics and Technology, Institute of Quantum Matter, South China Normal University, Guangzhou 510006, China}}

\newcommand{\scnt}{\affiliation{Southern Center for Nuclear-Science Theory (SCNT), Institute of Modern Physics, Chinese Academy of Sciences, Huizhou 516000, Guangdong Province, China}}

\newcommand{\shanghai}{\affiliation{School of Physics,  Faculty of Basic Sciences, University of Shanghai for Science and Technology, Shanghai 200093, China}}

\newcommand{\gscas}{\affiliation{Graduate School of China Academy of Engineering Physics, Beijing 100193, China}}

\begin{document}
\include{def-com}
\title{\boldmath Machine Learning Unveils Finite-volume Energy Shifts in Three-body System}

\author {Wei-Jie Zhang}
\iqm
\moe
\sfim
\author {Zhenyu Zhang} \email{Co-first author}
\iqm
\moe
\sfim
\ihep
\author {Jifeng Hu}
\email{hujf@m.scnu.edu.cn}
\iqm
\sfim
\author {Bing-Nan Lu}
\email{bnlv@gscaep.ac.cn}
\gscas
\author {Jin-Yi Pang}
\email{jypang@usst.edu.cn}
\shanghai
\author {Qian Wang}
\email{qianwang@m.scnu.edu.cn}
\iqm
\sfim
\scnt
 
\date{\today}

\begin{abstract}
Finite-volume extrapolation (FVE) is essential for extracting physical observables in the lattice calculation. While rigorous FVE formulations are well established for short-range potentials in both two- and three-body systems, long-range interactions with force ranges comparable to the lattice size $L$ remain challenging. Extending a previous data-driven scheme for two-body systems, we apply symbolic regression (PySR) to uncover universal three-body FVE formulae. For short-range potentials, we reproduce the two limiting cases, i.e. $\kappa_3\gg\kappa_2$ and $\kappa_3\sim\kappa_2$. For pure long-range potentials, we obtain a dedicated analytic expression, and after incorporating short-range contributions, we uncover a unified formula consistent with the original PySR solution, which performs excellently in the intermediate force range around 1 fm. This work demonstrates that combining machine learning with physical constraints can yield novel analytical results inaccessible to conventional theoretical tools, advancing data-driven methodologies in hadron physics.

\end{abstract}
\maketitle

\section{Introduction}
Lattice effective field theory (LEFT) provides an alternative framework in which nucleons, rather than quarks and gluons, are placed on a discretized spacetime lattice. The interactions between nucleons are organized according to chiral effective theory, and ground-state properties are computed via Auxiliary-Field Quantum Monte Carlo simulations~\cite{Lahde:2019npb}. A central challenge in LEFT, as in any other lattice calculations, is finite-volume effects. For few-body systems, the L\"uscher formalism and its extensions to composite particles~\cite{Bour:2011ef} provide analytic control over volume dependence, with topological corrections that depend only on the mass ratios of the constituents. For medium-mass nuclei, systematic Euclidean time extrapolation (``triangulation") using multiple trial states enables reliable extraction of infinite-volume observables~\cite{Lahde:2013uqa,Lahde:2014sla}. More recently, symmetry sign extrapolation has extended these methods to neutron-rich systems~\cite{Lahde:2015ona}. Volume extrapolation in LEFT corrects for the finite-volume contamination, ensuring that the observables converge to their infinite-volume limits.

As the result, finite-volume extrapolation (FVE) is the bridge combining the results from finite volume and infinite volume and it can also extract some physical observables. In two-body system, FVE for short-range potential case has been derived by L\"uscher formula~\cite{Luscher:1985dn}.  However, the long-range interactions, in which force range is comparable to the box size and light particles serve as mediators in scattering states, present non-negligible challenges caused by the changeable force range and the significant partial wave mixing~\cite{Bubna:2025gsd}. A typical example is the light-meson exchange in $DD^*$ system~\cite{Meng:2023bmz}. Two methods struggle to explore the law of FVE, one is modifying L\"uscher formula numerically by rewriting the zeta function and separating short-range and long-range potential~\cite{Bubna:2024izx,Dawid:2024dgy,Romero-Lopez:2020rdq,Bubna:2025gsd,Iritani:2018vfn,Yu:2025gzg},
the other is handling long-range interactions within the Hamiltonian framework and constructing irreducible representations and solve Lippmann-Schwinger equation to extract phase shifts~\cite{Meng:2021uhz}.
However, neither approach yields a general formula. In three-body system, studies for short-range potential bound states have been thoroughly studied~\cite{Meissner:2014dea,Hammer:2017uqm,Hansen:2016ync,Luscher:1985dn,Konig:2017krd,Doring:2018xxx} and present the formula in unitary limit by solving Schr\"odinger equation~\cite{Meissner:2014dea}, or adopt effective field theory based on particle-dimer picture to derive FVE in two extreme situations~\cite{Doring:2018xxx,Hammer:2017uqm}. For example, there exists a $DD^*K$ bound state using effective field theory approach, a short-range repulsive  three-body force is included~\cite{Zhang:2024yfj}. Nevertheless, unlike two-body system, FVE in three-body system for long-range potential faces is a more complicated situation due to the disruption of the invariance of the Efimov continuous scale in finite volume~\cite{Kreuzer:2013oya}.  

Symbolic regression (SR)~\cite{GeneticProgramming} can be applied to solve these problems; its advantages include the ability to uncover mathematical relationships from data~\cite{science.1165893} which has already exhibited its strengths in physics field~\cite{amil2009statistical}. Based on the genetic programming~\cite{Angelis2023} a vast space containing variables, constants and signs can be searched and finally a new formula is constructed~\cite{Cranmer:2020wew}. Compared with the traditional SR, some mathematics operations,i.e., symmetry, additive or multiplicative separability, variable substitution and modularity are utilized to add the physics limit and forge to present more effective results. For example, PhyE2E algorithm successfully improves the solar activity formula and discovers new analytical expressions for star formation rates from the simulation of the FIRE-2 galaxy~\cite{Ying2025}. AI Feynman algorithm combing with neural networks, large language model, deep learning and some necessary physics limits, perfectly derived more than 100 equations in \textit{Feynman Lectures on Physics}~\cite{Udrescu:2019mnk,udrescu2020aifeynman20paretooptimal}. In fact, its capability of generating mathematical expression has been widely used and accelerated the researches in high-energy physics~\cite{tenachi2023deep,Dong:2022trn}. For instance, model building and background fitting for data processing in experiments~\cite{Tsoi:2024pbn,Butter:2021rvz,Mengel:2023mnw}. SR prompts rediscovery of dual relation of Kawai--Lewellen--Tye and Bern--Carrasco--Johansson in scattering amplitudes~\cite{Moynihan:2026mbz}. Fragmentation functions extracted from data are good candidates for global fitting~\cite{Makke:2025zoy}. Besides, SR plays a vital role in exploring Beyond the Standard Model (BSM)   physics~\cite{AbdusSalam:2025the,AbdusSalam:2024obf},  unveiling deeper relationships like constants of standard model~\cite{Chekanov:2025wzw} and Feynman integral singularities~\cite{Liu:2025tje}, exploring unknown phenomena like Tsallis distribution~\cite{Makke:2024whm}. 

FVE formula for short-range potential in two-body system has already reproduced by neural network~\cite{Lu:2022joy}. We have already rediscovered the FVE formula derived by L\"uscher formula and have found a general formula in both short-range and long-range potentials by SR in two-body system~\cite{Zhang:2025juc}. The similar method is utilized in three-body system in our research. Before applying SR, samples  need to prepared. Hadron Lattice Effective Field Theory (HLEFT)~\cite{Zhang:2024yfj}, which is analogous to Nuclear Lattice Effective Field Theory (NLEFT)~\cite{Meissner:2022cbi,Lee:2025req,Lee:2008fa}, a powerful method for \textit{ab initio} nuclear structure calculation~\cite{Lee:2008fa} combining Chiral Effective Field Theory and Lattice quantum Monte Carlo (QMC)~\cite{Lahde:2019npb,Lahde:2013kma,Lahde:2013png} to solve many-body problems. Continuous spacetime is discretized on a lattice where nucleons are placed~\cite{Lee:2025req} and the interaction between nucleus originates from Chiral Perturbation Theory. 
\begin{figure}[htbp!]
    \centering
    \includegraphics[width=1.1\linewidth]{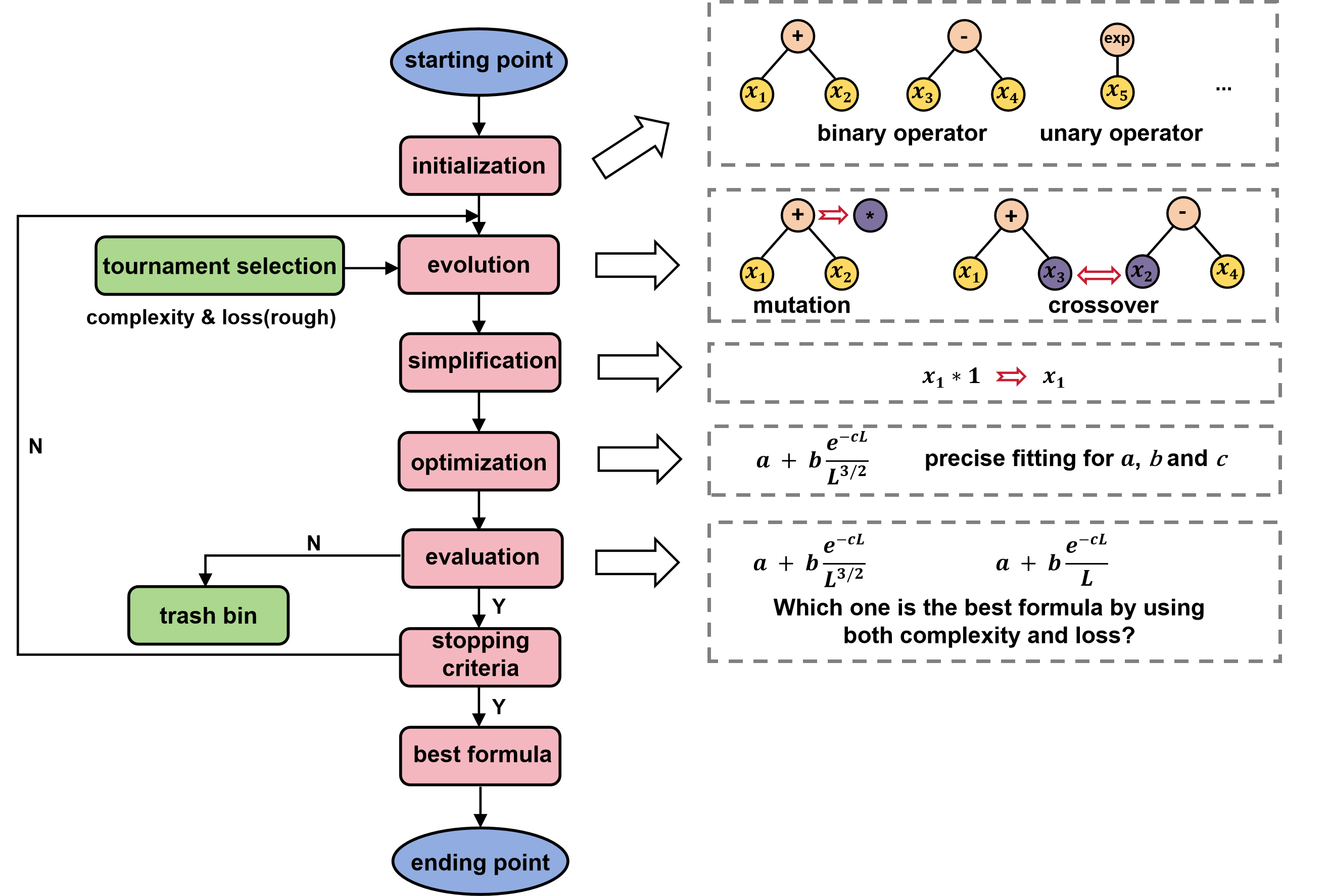}
    \caption{Each evolution process in PySR. The baby blue oval frames represent the starting point and ending point of evolution. The light pink and light shallow green rectangular frames represent the process in one iteration. The gray dashed rectangular frames are the details for the evolution process of light pink rectangular frames.}
    \label{fig:PySR_Each_Evolution_Process}
\end{figure}
NLEFT has already demonstrated its irreplaceable role in exploring structures and interactions in nucleus, especially in some characteristics of nucleon like its ground states~\cite{Borasoy:2006qn,Epelbaum:2009pd,Epelbaum:2010xt,Lahde:2013uqa,Lu:2018bat,Lu:2021tab,Elhatisari:2022zrb} and excited states~\cite{Epelbaum:2013paa,Shen:2022bak,Meissner:2023cvo,Shen:2024qzi}, distributions of nuclear intrinsic density and clustering phenomena~\cite{Epelbaum:2011md,Epelbaum:2012qn,Epelbaum:2012iu,Elhatisari:2017eno,Zhang:2024wfd}, scattering of nucleus-nucleus~\cite{Bour:2012hn,Elhatisari:2015iga}, investigation of both zero-temperature and finite-temperature nuclear matter properties~\cite{Elhatisari:2016owd,Lu:2019nbg,Ren:2023ued,Ma:2023ahg} calculation of $\alpha$-cluster structure and light nuclei~\cite{Lahde:2019npb,Lahde:2013png,Bao-Ge:2024djr,Alarcon:2015vds,Lahde:2013kma} explanation of exotic quantum phase transition~\cite{Lahde:2019npb,Agar:2026nxr} and structure of exotic nucleon~\cite{Haidenbauer:2025zrr,Hildenbrand:2024ypw}. 
The researches have also been expanded to hypernuclear systems~\cite{Zhang:2024yfj,Bour:2014bxa,Scarduelli:2020xae,Hildenbrand:2024ypw}. However, FVE challenges still exist in nucleon system.

In Ref.~\cite{Zhang:2025juc}, we have already discussed FVE formula using SR in two-body identical bosons system, and SR provides the result combining short-range and long-range potential. Following the approach in two-body system, in this work, we introduce SR approach to study FVE formula in three-body identical bosons system, our workflow can be divided into these parts below. In Sec.~\ref{sec:Samples}, we introduce the samples, generated in limited box sizes,  for both short-range and long-range potential cases. The details of SR are presented in Sec.~\ref{sec:Symbolic}. In Sec.~\ref{sec:Results}, we describe the discovery of three-body FVE formulae applicable to both short-range and long-range potentials, deriving a unified formula.

\begin{figure*}[htbp]
\centering
    \includegraphics[width=0.44\textwidth]{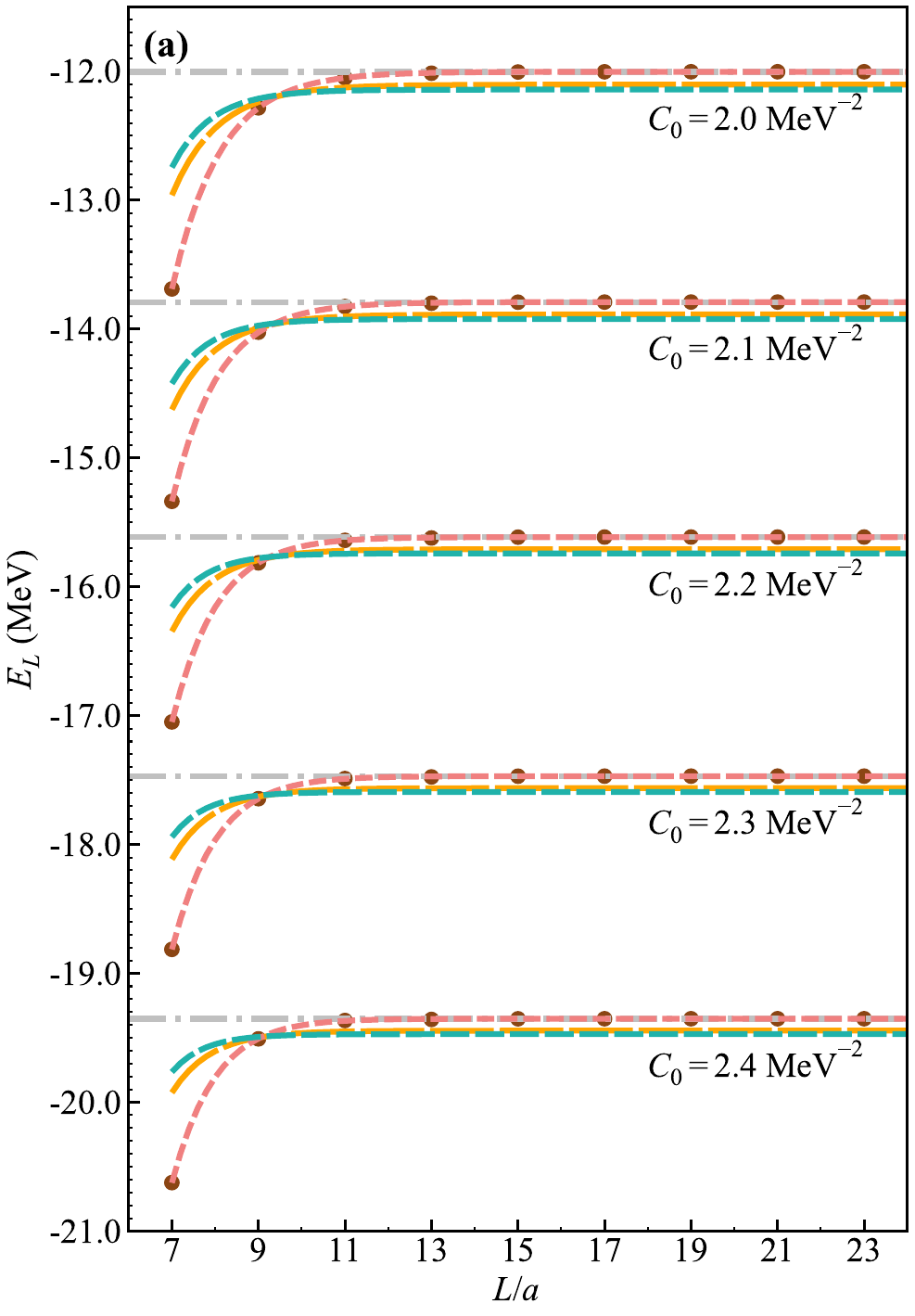}\includegraphics[width=0.44\textwidth]{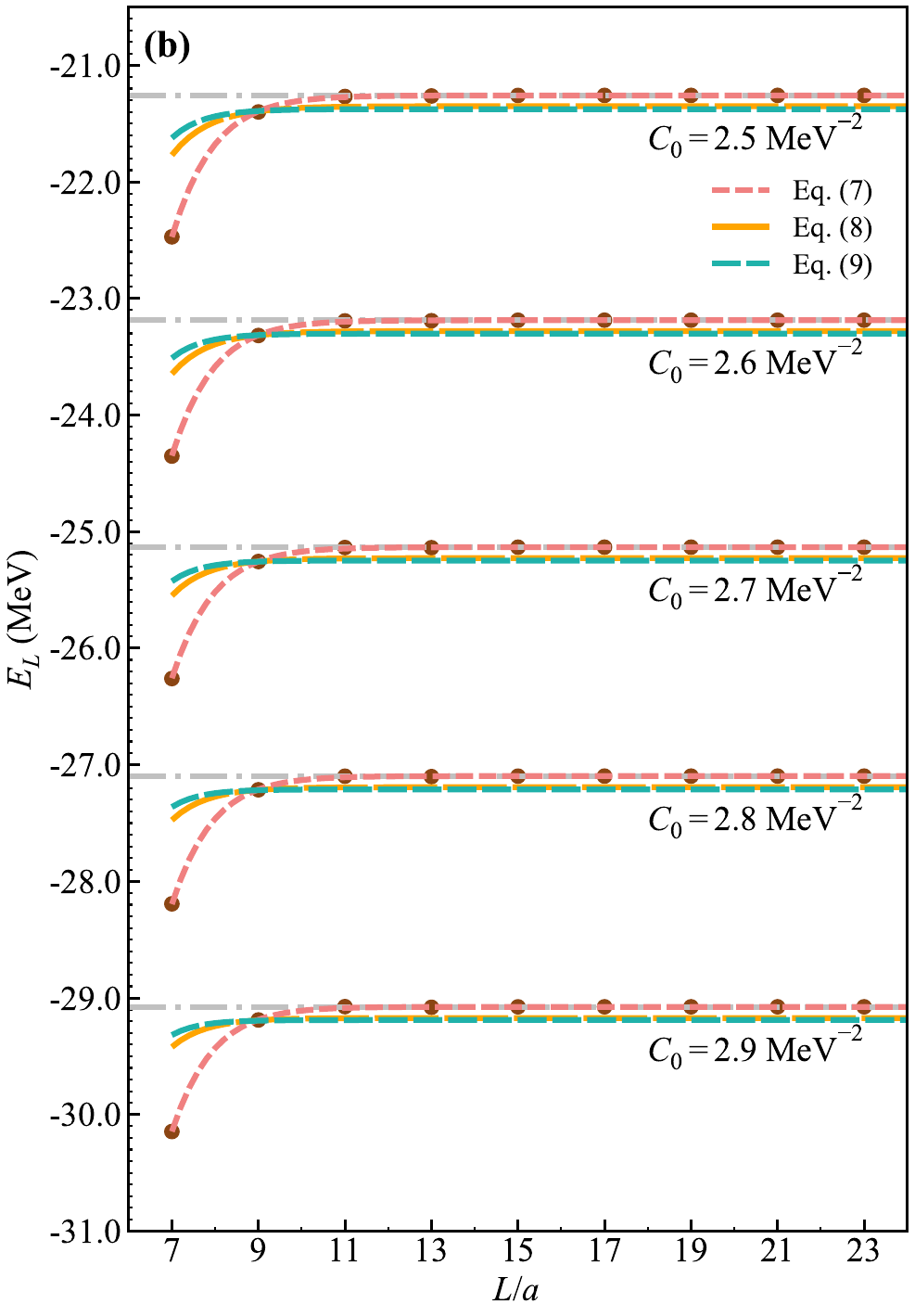}
    \caption{The fitting results for short-range case. The brown points are the samples generated by LEFT of short-range potential, corresponding to ten potential strengths $C_0$ =2.0, 2.1, $\cdots$, 2.9 MeV$^{-2}$. Pink dashed, orange solid and seagreen long-dashed lines corresponding to Eq.~\eqref{Eq: PySR_3b_Short_Range_Formula}, Eq.~\eqref{Eq: UL_3b_Short_Range_Formula} and Eq.~\eqref{Eq: 2b_Short_Range_Formula}. The gray dot-dashed line is the parameter $A_1$ in Eq.~\eqref{Eq: PySR_3b_Short_Range_Formula}. The box size $L^3 = 7,9,\cdots, 23 $ fm$^3$.} 
    \label{Fig:PySR_FVE_Short_Range}
\end{figure*}
\section{Samples Generation}\label{sec:Samples}
For a simplicity, we consider a system of three identical spinless particles under a non-relativisitc framework. All the particles masses are set to $m = 1969 ~\mathrm{MeV}$. 
Considering only the pairwise interactions, the short-range potential in coordinate space takes the form:
\begin{equation}
    V(\boldsymbol{r}_{ij})=-C_0\delta^3(\boldsymbol{r}_{ij}),
    \label{Eq:short_V}
\end{equation}
where $\boldsymbol{r}_{ij} = \boldsymbol{r}_i - \boldsymbol{r}_j$ is the relative coordinate between the $i$th and $j$th particles in the center-of-mass frame. $C_0$ represents the strength of potential. From the effective field theory point of view, for the potential $V(\boldsymbol{r}_{ij})$, a physical cutoff is needed to distinguish the short-range contribution from the long-range one, which keeps the renormalization group invariant. Here a single-particle regulator~\cite{Lu:2023jyz,Entem:2003ft,Epelbaum:2004fk} $f(\boldsymbol{p}^{(\prime)}_i,\boldsymbol{p}^{(\prime)}_j)=\prod
\limits_{i=1}^2g_{\Lambda}(\boldsymbol{p}_i)g_{\Lambda}(\boldsymbol{p}^\prime_i)$ can be utilized, where $g_{\Lambda}(\boldsymbol{p})=$exp$(-\boldsymbol{p}^6/2\Lambda^{6})$~\cite{Wang:2026xyn} is a soft cutoff function with $\boldsymbol{p}_i$ and $\boldsymbol{p}^\prime_i$ the momenta of the incoming and outgoing of the $i$th particle~\cite{Zhang:2024yfj}. Thus the Hamiltonian can be written as:
\begin{equation}
H=\sum_{i=1}^3\frac{\boldsymbol{p}_i^2}{2m_i}+\sum_{i<j}^3f(\boldsymbol{p}^{(\prime)}_i,\boldsymbol{p}^{(\prime)}_j)V(\boldsymbol{q}),
\label{Eq:H}
\end{equation}
where $m_i$ is the mass of the $i$th particle, $\bm{q}=\bm{p}-\bm{p}'$ is the transferred momentum of the two particles and $\boldsymbol{p}$ ($\boldsymbol{p}^\prime$) represents the relative incoming (outgoing) momentum between $\boldsymbol{p}_i$ ($\boldsymbol{p}^{\prime}_i$) and $\boldsymbol{p}_j$ ($\boldsymbol{p}^{\prime}_j$). The corresponding potential in momentum space $V(\boldsymbol{q})$ = $-C_0$ can be obtained by a Fourier transformation from the expression in coordinate space. 

In our framework, HLEFT~\cite{Zhang:2024yfj} is used to generate the samples for both short-range and long-range potentials. 
The solution of the Schr\"{o}dinger equation can be obtained by directly diagonalizing the Hamiltonian defined on lattice coordinate space. The matrix exact diagonalization scheme borrows the implicitly restarted Lanczos method, which finds the eigenvalues and eigenvectors~\cite{lehoucq1998} by SciPy~\cite{Virtanen:2019joe} in Python. In short-range potential case, we perform simulations on cubic lattice $L^3 = 7^3, 8^3, \cdots,24^3 ~\mathrm{fm}^3$. The potential strength $C_0$ has various values to generate the samples. More details of these samples are listed in Appendix~\ref{app:Samples}.
\begin{figure}[htbp]
    \centering
    \includegraphics[width=0.96\linewidth]{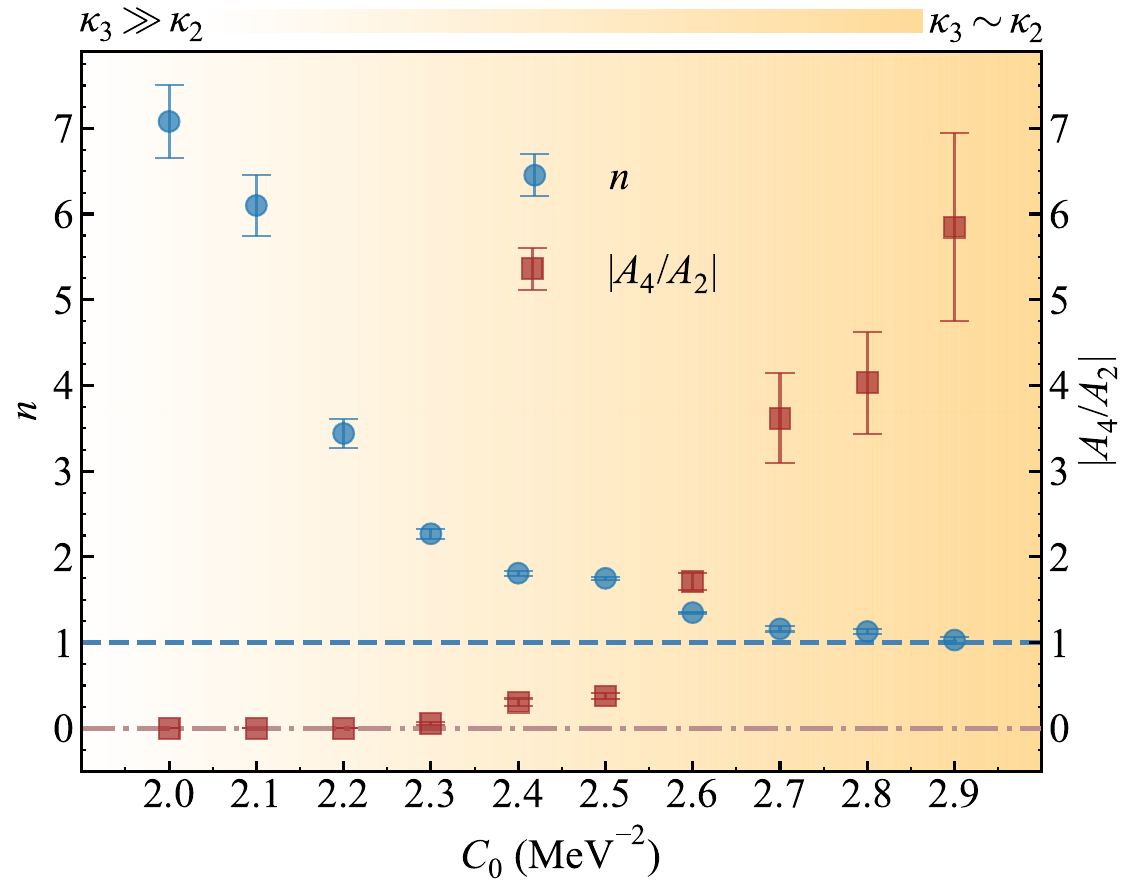}
    \caption{The fitting results of parameters the power $n$ of $L$ and the important ratio $|A_4/A_2|$ for all samples with short-range potential. The blue dots and the red squares represent the parameters $n$ and $|A_4/A_2|$ in Eq.~\eqref{Eq: PySR_3b_Short_Range_Formula}, respectively. The blue dashed line is asymptotic line $n = 1$, and the red dot-dashed line is asymptotic line $|A_4/A_2|=0$. The background color approaches orange means $\kappa_3\sim\kappa_2$.}
    \label{Fig:Short_Range_n_and_C4C2_Result}
\end{figure}

In the case of long-range potential, in each pair of particles, we define the potential like Yukawa potential form~\cite{Bubna:2024izx},
\begin{align}
    V(\bm{r}_{ij}) = -C_{01}\delta^3(\bm{r}_{ij})-C_{02}\frac{e^{-\mu r_{ij}}}{r_{ij}},
    \label{Eq:long_V}
\end{align}
where $\mu$ reflects the range of the force. $C_{01}$ and $C_{02}$ are the strengths of the short-range potential and long-range potential. For simplicity, we set $C_{01} = C_{02}$ when generating samples. The details of the ranges of the force and $C_{01} = C_{02}$ values can be found in Appendix~\ref{app:Samples}. The reason of the existence of a short-range potential term in Eq.~\eqref{Eq:long_V} is due to the divergence of the Yukawa potential in $r_{ij}\to 0$, which means that it needs a regularization. Similarly to the short-range potential case, a reasonable cutoff is applied, which gives reasonable physical observables. For the short-range potential term, $f(\boldsymbol{p}^{(\prime)}_i,\boldsymbol{p}^{(\prime)}_j)$ is still applied to each pair of particles. At the same time, the regulator of the long-range potential~\footnote{ Here, $\Lambda$ is the cutoff to regularize the long-range potential, which is set equal to that for the short-range one above.} is $\hat{f}(\boldsymbol{q})$ = exp$[-(\boldsymbol{q}^2+\mu^2)/\Lambda^2]$~\cite{Meng:2023bmz,Reinert:2017usi}. With the definition of the regulators, the Hamiltonian can be written as:
\begin{equation}
\begin{split}
 H=\sum_{i=1}^3\frac{\boldsymbol{p}_i^2}{2m_i}&+\sum_{i<j}^{3}f(\boldsymbol{p}^{(\prime)}_i,\boldsymbol{p}^{(\prime)}_j)V_S(\boldsymbol{q})+\sum_{i<j}^{3}\hat{f}(\boldsymbol{q})V_L(\boldsymbol{q}).   
\end{split}
\label{Eq:long_H}
\end{equation}
Here, both the short-range potential $V_S(\boldsymbol{q}) = -C_{01}$ and the long-range potential $V_L(\boldsymbol{q}) \sim \frac{1}{\bm{q}^2+\mu^2}$ can be transformed into their coordinate forms by Fourier transformation. The simulations are performed in cubic lattice of box sizes $L^3 = 10^3,11^3,\cdots,24^3$ fm$^3$. Finally, analogous to our treatment for the pure short-range potential case, we solve the Schr\"{o}dinger equation and extract the finite-volume binding energy $E_L$ via the HLEFT. Representative examples are provided in Appendix~\ref{app:Samples}.

\section{Symbolic Regression}\label{sec:Symbolic}

Samples generated by HLEFT are performed for SR to explore FVE formula. SR model based on genetic algorithms (GA)~\cite{Ireland:2004kp,Fernandez-Ramirez:2008ixe} and can search for the mathematical relationship between inputs and outputs through multi-population evolutionary~\cite{Cranmer:2020wew}. PySR model~\cite{Cranmer:2023pysr} explores the analytic expression space constructing by the sets of input(output) variables, constant terms and operators. The input variable here is box size $L$ and the output variable is binding energy $E_L$. The operators contain addition, multiplication, division, exponential, sqrt, cube and some custom-defined operators. Compared with the massive samples utilized in deep learning, few samples are sufficient for PySR. The model performs 200 iterations of 200 different population samples, with each population containing 100 individuals. The evaluation of the output formula contains two vital elements: loss and complexity. Loss is used to measure how well the output formula describes the samples, which is defined as mean square error (MSE):
 \begin{equation}
\mathrm{Loss}=\sum_{i=1}^{N} (E_{\mathrm{PySR}}(L_{i}) - E_L({L_{i}}))^2/N.
\label{Eq:loss}
\end{equation}
Here, $E_{\mathrm{PySR}}$ is the formula output by the PySR model, and $E_L$ is the energy of a system in a box of size $L^3$ calculated by HLEFT. Complexity is an indicator for measuring the degree of redundancy in a formula. Values of complexity for each variable, constant and operator can be set. PySR model combines these two elements and evaluates the formula using score, which rewards minimal loss and penalizes more complicated formula. It can be defined as
\begin{equation}
\mathrm{Score}\equiv-\frac{\Delta\ \mathrm{ln(Loss)} }{\Delta\ C}.
\label{Eq:score}
\end{equation}
Here, $C$ is the complexity~\footnote{The upper limit of the $C$ value can be set larger to allow higher order contributions as that in analytic analysis ~\cite{Luscher:1985dn}.} (each complexity of variables, constants and operators can be found in the Appendix~\ref{app:Introduction}), which is defined as the total number of operations, variables, and constants used in a formula. Score can help PySR choose the best formula.

The whole evolution process in PySR is similar to the process in nature. 200 populations are included in evolutions and each of them has their own mechanism. For each evolution, initial formulae are randomly constructed by variables, constants and operators and experience mutations and crossovers in each iteration. PySR chooses some performed-well individuals as the parents for next iteration until triggering the stop condition set before. Each evolution process can be seen in Fig.~\ref{fig:PySR_Each_Evolution_Process}. Different evolutions happen at the same time and populations exchange some elements and the best formula can be chosen through this process.

\section{Results}\label{sec:Results}

\begin{figure*}[thbp]
\centering
    \includegraphics[width=0.32\textwidth]{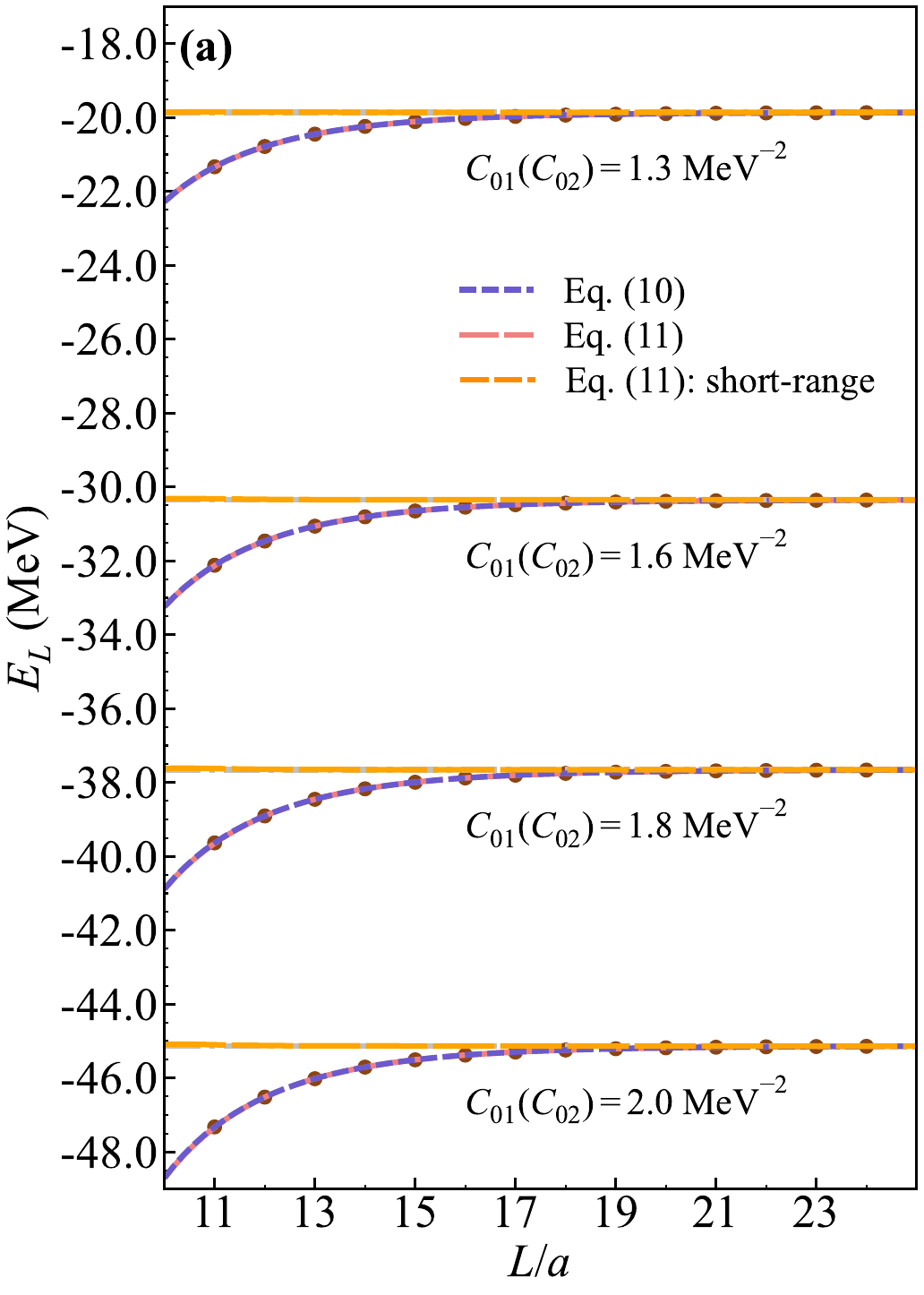}
    \includegraphics[width=0.32\textwidth]{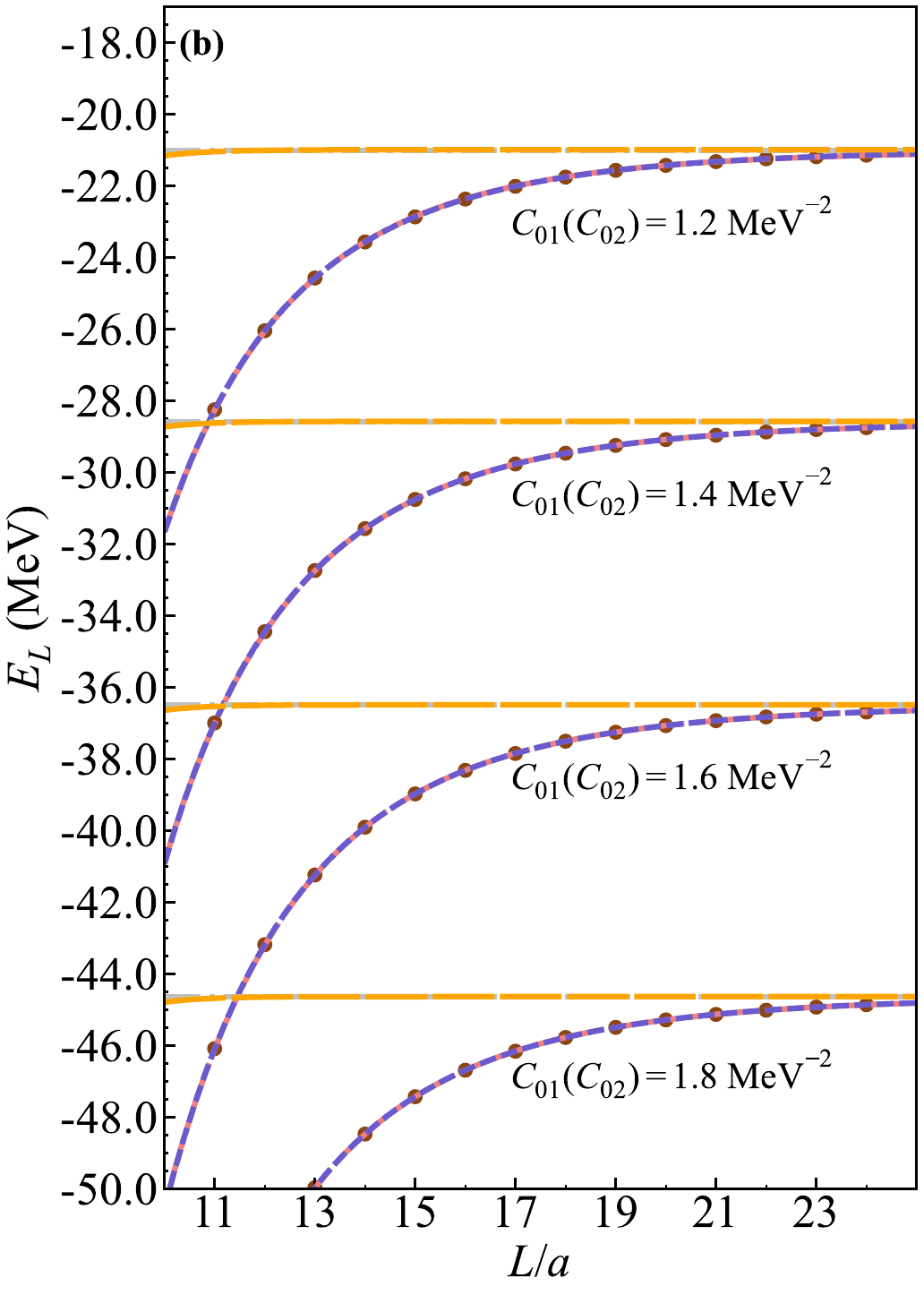}
    \includegraphics[width=0.32\textwidth]{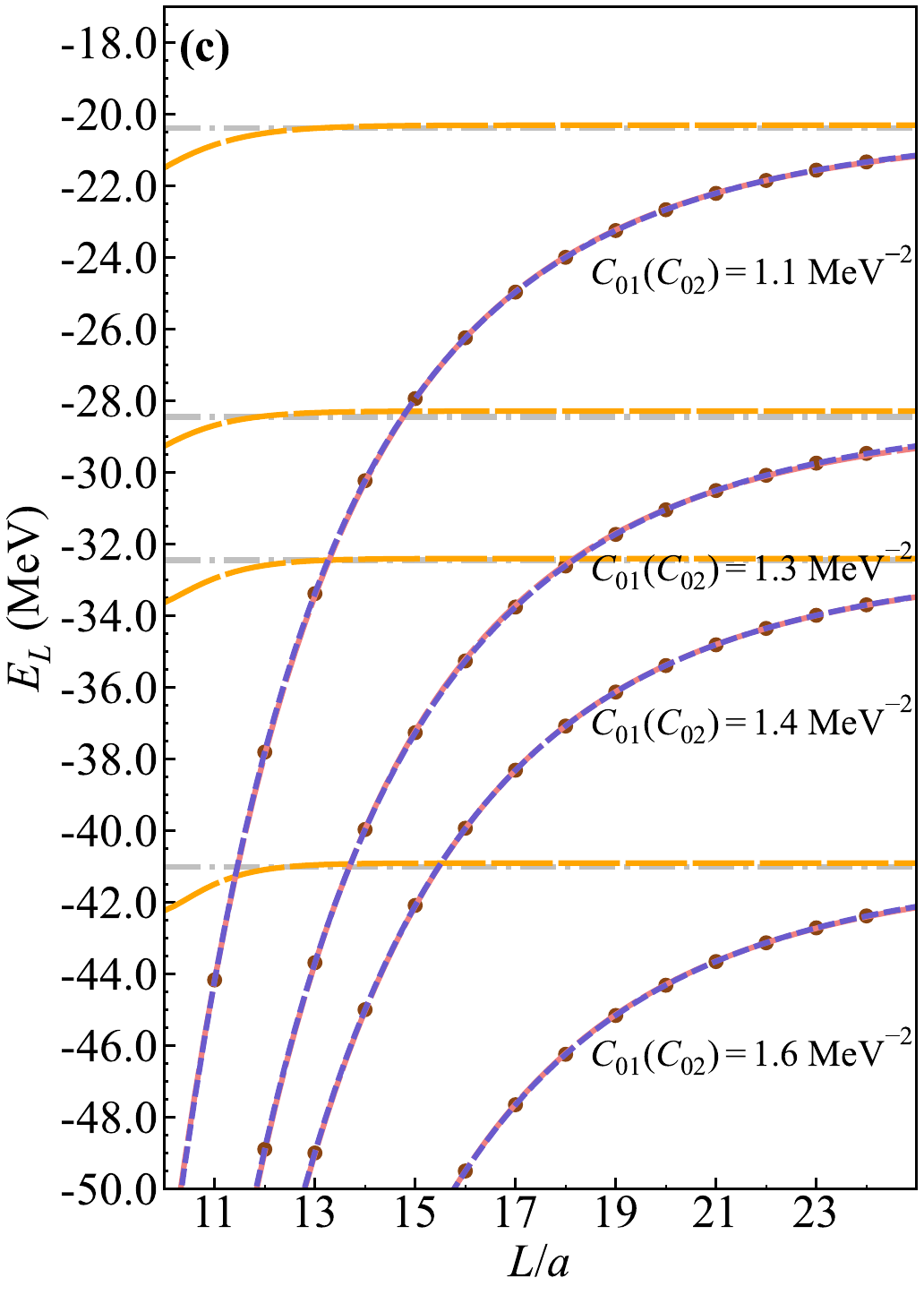}
    \caption{The fitting results for long-range case. The pictures (a), (b), and (c) are the fitting results for 3, 5, and 8 fm samples, respectively. 
    The brown points are the samples generated by LEFT of long-range potential, corresponding to different potential strengths $C_{01}(C_{02})$. Purple, pink, and orange dashed lines corresponding to Eq.~\eqref{Eq: PySR_3b_Long_Range_Formula}, Eq.~\eqref{Eq: PySR_3b_Short_Long_Range_Formula}, and the contribution of the short-range potential in Eq.~\eqref{Eq: PySR_3b_Short_Long_Range_Formula}. The gray dashed line is the fitting value of $\mathcal{A}_1$ in Eq.~\eqref{Eq: PySR_3b_Long_Range_Formula}. The box size $L^3 = 10^3, 11^3 \cdots24^3$ fm$^3$.} 
    \label{Fig:PySR_FVE_Long_range}
\end{figure*}

As discussed in the previous section, we consider three identical spinless bosons with mass $m=1969~\mathrm{MeV}$, forming a bound state with binding momentum $\kappa_3$, interacting via two-body forces only. 
The generated samples for short-range potential case input to PySR model, the best result presented by PySR is:
\begin{equation}
    E_L = A_1 + \frac{A_2}{L^{3/2}}e^{-A_3L}+\frac{A_4}{L^{n}}e^{-A_3L}.
    \label{Eq: PySR_3b_Short_Range_Formula}
\end{equation}
Here $A_{1,2,3,4}$ are free parameters, with $A_{1}$ representing the binding energy in the infinite-volume limit, $A_{2}$ and $A_{4}$ characterizing to the wave-function normalization coefficient, and $A_{3}$ scaling with the binding momentum.
In fact, in short-range potential case, theoretical research of FVE formula has already reached conclusions~\cite{Meissner:2014dea,Doring:2018xxx}. FVE formula depends on both two-body binding momentum $\kappa_2$ and three-body binding momentum $\kappa_3$. Under unitary limit ($\kappa_3 \gg \kappa_2$), theoretical formula is obtained based on L\"ushcer method in Ref.~\cite{Luscher:1985dn}. By solving the three-body Schr\"odinger equation in a finite cubic box and employing the asymptotic form of the infinite-volume three-body bound state wave function, the exponential decay behavior of the energy correction is obtained~\cite{Meissner:2014dea}.
\begin{equation}
    E_L = E_\infty + \frac{A}{L^{3/2}}e^{-\frac{2}{\sqrt{3}}\kappa_3L}.
    \label{Eq: UL_3b_Short_Range_Formula}
\end{equation}
Here, $E_\infty$ represents the binding energy in infinite volume. $A$ is the free parameter. For the case each pair of particles is tightly bound, i.e. $\kappa_3 \sim \kappa_2$, formula~\cite{Doring:2018xxx} in short-range potential is governed by:
\begin{equation}
    E_L =E_\infty + \frac{A}{L}e^{-\frac{2}{\sqrt{3}}\sqrt{\kappa_3^2-\kappa_2^2}L}.
    \label{Eq: 2b_Short_Range_Formula}
\end{equation}
To compare the performance of different formulae against the samples, we present the fitting results in Fig.~\ref{Fig:PySR_FVE_Short_Range} for the short-range potential using three candidate expressions: Eq.~\eqref{Eq: PySR_3b_Short_Range_Formula} (pink dashed line), Eq.~\eqref{Eq: UL_3b_Short_Range_Formula} (orange dashed solid line), and Eq.~\eqref{Eq: 2b_Short_Range_Formula} (seagreen long-dashed line). The gray dot-dashed line is the fitting result of $A_1$ in Eq.~\eqref{Eq: PySR_3b_Short_Range_Formula}, which represents $E_\infty$. For all samples of different strength of potential $C_0$, Eq.~\eqref{Eq: PySR_3b_Short_Range_Formula} provides the best description of these samples across the full ranges of box sizes compared with Eq.~\eqref{Eq: UL_3b_Short_Range_Formula} and Eq.~\eqref{Eq: 2b_Short_Range_Formula}.

\begin{figure}[htbp]
    \centering
    \includegraphics[width=1.0\linewidth]{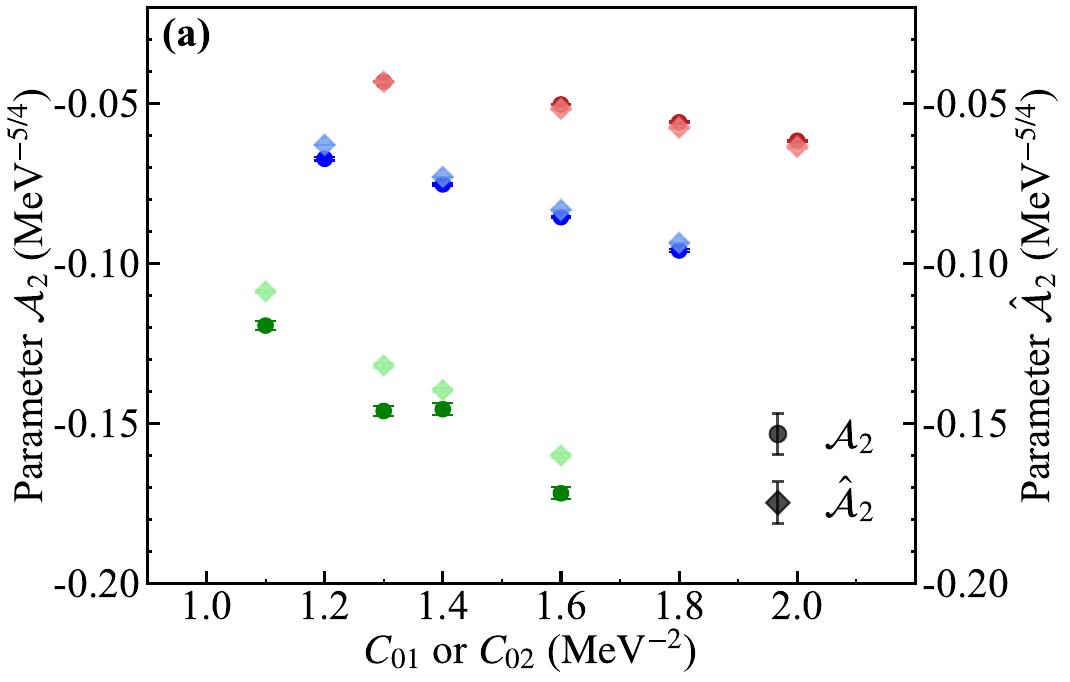}
    \includegraphics[width=1.0\linewidth]{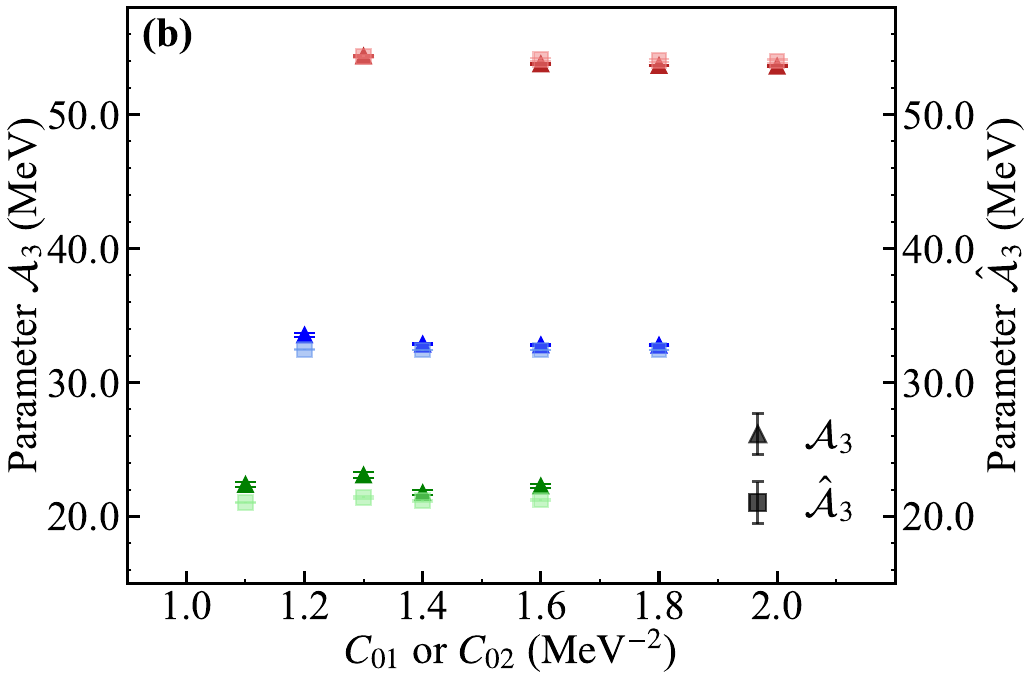}
    \caption{The fitting results of the parameters $\mathcal{A}_2$, $\mathcal{A}_3$, $\hat{\mathcal{A}}_2$ and 
    $\hat{\mathcal{A}}_3$ at various $C_{01}(C_{02})$s. The dots(triangles) and diamonds(squares) represent the parameters $\mathcal{A}_2$($\mathcal{A}_3$) and $\hat{\mathcal{A}}_2$($\hat{\mathcal{A}}_3$), respectively. The colors red, blue and green are used to distinguish the results in force ranges 3, 5, and 8 fm, respectively. }
    \label{Fig:Long_Range_D2E4_D3E5}
\end{figure}

Taking it a step further, we analyze the properties of the parameters in Eq.~\eqref{Eq: PySR_3b_Short_Range_Formula}, especially the power $n$ of $L$ in the third term, the important ratio $|A_4/A_2|$ between the third term compared and the second term. $n$ and $|A_4/A_2|$ values with different potential strengths $C_0$ are shown in Fig.~\ref{Fig:Short_Range_n_and_C4C2_Result}. Here, the blue dots and the red squares represent the parameters $n$ and $|A_4/A_2|$ in Eq.~\eqref{Eq: PySR_3b_Short_Range_Formula}, respectively.  
It can be seen that when the potential strengths $C_0$ increases from 2.0 to 2.9 $\mathrm{MeV}^{-2}$, parameter $n$ gradually decreases and converges to 1, which approaches to that in Eq.~\eqref{Eq: 2b_Short_Range_Formula} as $\kappa_3\sim \kappa_2$. In contrast, when the potential strengths $C_0$ decrease from 2.9 to 2.0 $\mathrm{MeV}^{-2}$, ratio $|A_4/A_2|$ gradually decreases and converges to 0, which recover the formula in Eq.~\eqref{Eq: UL_3b_Short_Range_Formula}, i.e. the case of $\kappa_3\gg \kappa_2$. It should be noted that, $C_0$ is strongly correlated with the momentum ratio $\kappa_3/\kappa_2$. In Fig.~\ref{Fig:Short_Range_n_and_C4C2_Result}, the region where the color approaches colorless indicates that $\kappa_3 \gg \kappa_2$ and approaches orange means $\kappa_3 \sim \kappa_2$. A smaller value of $C_0$ indicates the more loosely bound states for three-body system, whereas a larger value of $C_0$ represents more tightly binding for two-body subsystem. 
Thus, when three particles bind as a loosely bound state, i.e., $\kappa_3\gg\kappa_2$, the second term of the Eq.~\eqref{Eq: PySR_3b_Short_Range_Formula} plays the main role,  i.e., $\Delta E=\frac{A_2}{L^{3/2}}e^{-A_3L}$, and if two-body subsystem tightly binds together,  i.e., $\kappa_3\sim\kappa_2$, the third term of the formula plays the main role, i.e., $\Delta E=\frac{A_4}{L}e^{-A_3L}$, which has rediscovered the theoretical formula Eqs.~\eqref{Eq: UL_3b_Short_Range_Formula} and \eqref{Eq: 2b_Short_Range_Formula}. However, the LEFT data considered here do not reach either of the two extreme regimes, $\kappa_3\gg\kappa_2$ and $\kappa_3\sim\kappa_2$; consequently, the exponent $A_3$ is distinct from both the exponent in Eq.~\eqref{Eq: UL_3b_Short_Range_Formula} and that in Eq.~\eqref{Eq: 2b_Short_Range_Formula}. Nevertheless, within uncertainties, $A_3$ still lies between these two values. Finally, our results show that between these two extreme situations, there’s a linear combination of Eqs.~\eqref{Eq: UL_3b_Short_Range_Formula} and \eqref{Eq: 2b_Short_Range_Formula} in the intermediate region. All the results of data mentioned above are listed in Appendix~\ref{app:PySR}.

For long-range potential, we set force range parameter $\mu$ to three values, which correspond to the force ranges of about 3, 5, 8 fm, and then generate samples and feed them into PySR model. The best result presented by PySR is:
\begin{equation}
    E_L = \mathcal{A}_1 + \frac{\mathcal{A}_2}{L^{9/4}}e^{-\mathcal{A}_3L}.
    \label{Eq: PySR_3b_Long_Range_Formula}
\end{equation}
Here, $\mathcal{A}_{1,2,3}$ are free parameters. The fitting results for long-range potential samples using this formula are listed in Fig.~\ref{Fig:PySR_FVE_Long_range}, described by the purple dashed line, which matches well with the samples in both small box sizes and large box sizes. The gray dashed line is the fitting value of $\mathcal{A}_1$ in Eq.~\eqref{Eq: PySR_3b_Long_Range_Formula}, which represents $E_\infty$. Overall, the Eq.~\eqref{Eq: PySR_3b_Long_Range_Formula} provides a robust and physically meaningful description of the system.

However, the contribution of short-range potential in the FVE formula should not be ignored. Based on the theoretical formula~\cite{Meissner:2014dea,Doring:2018xxx} from short-range potential case, i.e., Eqs.~\eqref{Eq: UL_3b_Short_Range_Formula} and \eqref{Eq: 2b_Short_Range_Formula}, our formula is rewritten as,
\begin{equation}
    E_L = \hat{\mathcal{A}}_1+\frac{\hat{\mathcal{A}}_2}{L^{9/4}}e^{-\hat{\mathcal{A}}_3L}+\frac{\hat{\mathcal{A}}_4}{L^{3/2}}e^{-\frac{2}{\sqrt{3}}\kappa_3 L}+\frac{\hat{\mathcal{A}}_5}{L}e^{-\frac{2}{\sqrt{3}}\sqrt{\kappa_3^2-\kappa_2^2}L},
    \label{Eq: PySR_3b_Short_Long_Range_Formula}
\end{equation}
keeping the last two terms as the short-range expression.
Here, $\hat{\mathcal{A}}_{1,2,3,4,5}$ are free parameters, and $\hat{\mathcal{A}}_1$ represents $E_\infty$, the second term represents the contribution of the long-range, while the third and fourth terms represent the contribution of the short-range potential. The fitting results for the long-range potential samples are also listed in Fig.~\ref{Fig:PySR_FVE_Long_range}, with the pink dashed line representing Eq.~\eqref{Eq: PySR_3b_Short_Long_Range_Formula} and the orange line showing the contribution of the short-range potential in Eq.~\eqref{Eq: PySR_3b_Short_Long_Range_Formula}. For all samples of different strength of potential $C_{01}(C_{02})$, Eq.~\eqref{Eq: PySR_3b_Short_Long_Range_Formula} provides well description of these samples, and similar to Eq.~\eqref{Eq: PySR_3b_Long_Range_Formula}. It's worth mentioning that the contribution of the short-range potential in Eq.~\eqref{Eq: PySR_3b_Short_Long_Range_Formula} hardly affect the fitting results. 

Taking a further step, we analyze the properties of the parameters in Eq.~\eqref{Eq: PySR_3b_Long_Range_Formula} and Eq.~\eqref{Eq: PySR_3b_Short_Long_Range_Formula}, especially the parameters $\mathcal{A}_2$, $\hat{\mathcal{A}}_2$ and $\mathcal{A}_3$, $\hat{\mathcal{A}}_3$ in the contribution of long-range potential in the second term. The fitting results of the parameters for all samples with different potential strengths $C_{01}(C_{02})$ and force ranges are shown in Fig.~\ref{Fig:Long_Range_D2E4_D3E5}. 
Here, the dots(triangles) and diamonds(squares) represent the parameters $\mathcal{A}_2$($\mathcal{A}_3$) and $\hat{\mathcal{A}}_2$($\hat{\mathcal{A}}_3$) in Eq.~\eqref{Eq: PySR_3b_Long_Range_Formula} and Eq.~\eqref{Eq: PySR_3b_Short_Long_Range_Formula}, respectively. Fitting results of samples with force ranges of about 3, 5, and 8 fm, distinguished by red, blue, and green, respectively. It can be seen that the values of parameters $\mathcal{A}_2$ and $\hat{\mathcal{A}}_2$ are close to each other, and so are those of $\mathcal{A}_3$ and $\hat{\mathcal{A}}_3$. This shows that the contribution of the short-range potential in Eq.~\eqref{Eq: PySR_3b_Short_Long_Range_Formula} hardly affect the fitting results.
It's worth mentioning that, parameters $\mathcal{A}_2$ and $\hat{\mathcal{A}}_2$ correlate with both the range of the force and the strength of the potential. Particularly, their absolute values are positively correlated to the force range and the force strength. In contrast, $\mathcal{A}_3$ and $\hat{\mathcal{A}}_3$
depend almost entirely on the range of the force and are negatively corrected with it. 
The behavior of all the parameters indicates that the correction of FVE is positively correlated with the range of the force, which matches what we expect physically. It follows that, Eq.~\eqref{Eq: PySR_3b_Long_Range_Formula} is equivalent to  Eq.~\eqref{Eq: PySR_3b_Short_Long_Range_Formula} in long-range potential case, and PySR gives a simple FVE formula.

\begin{figure*}[htbp]
    \centering
    \includegraphics[width=0.4\linewidth]{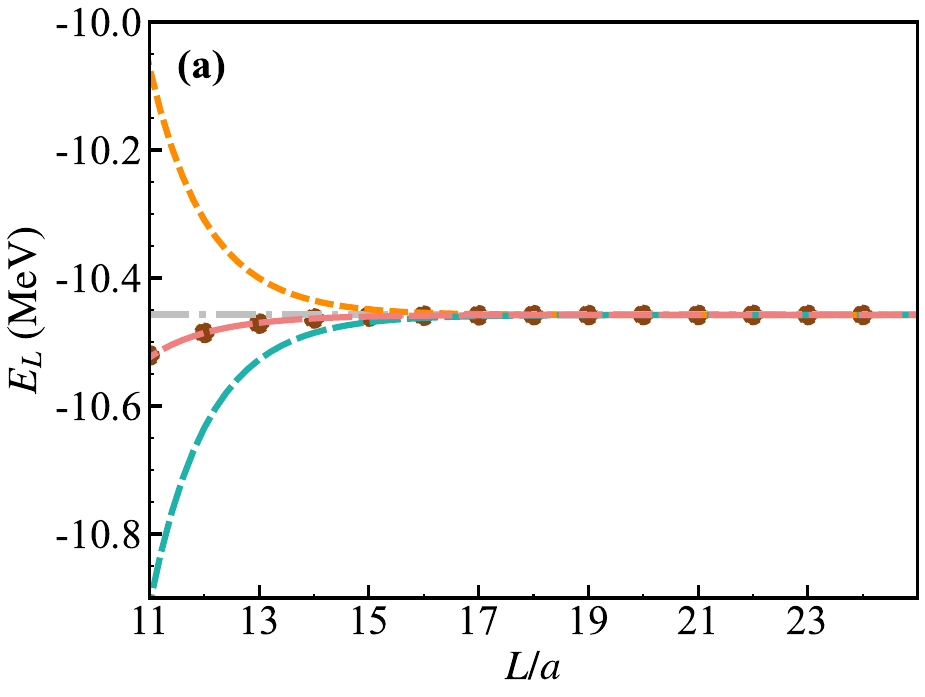}\includegraphics[width=0.4\linewidth]{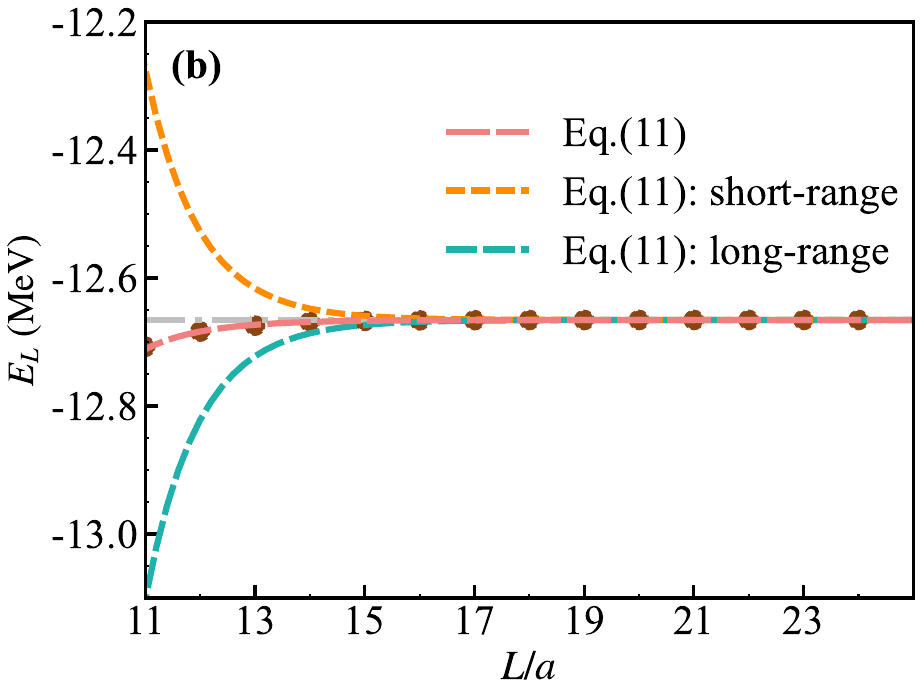}
    \includegraphics[width=0.4\linewidth]{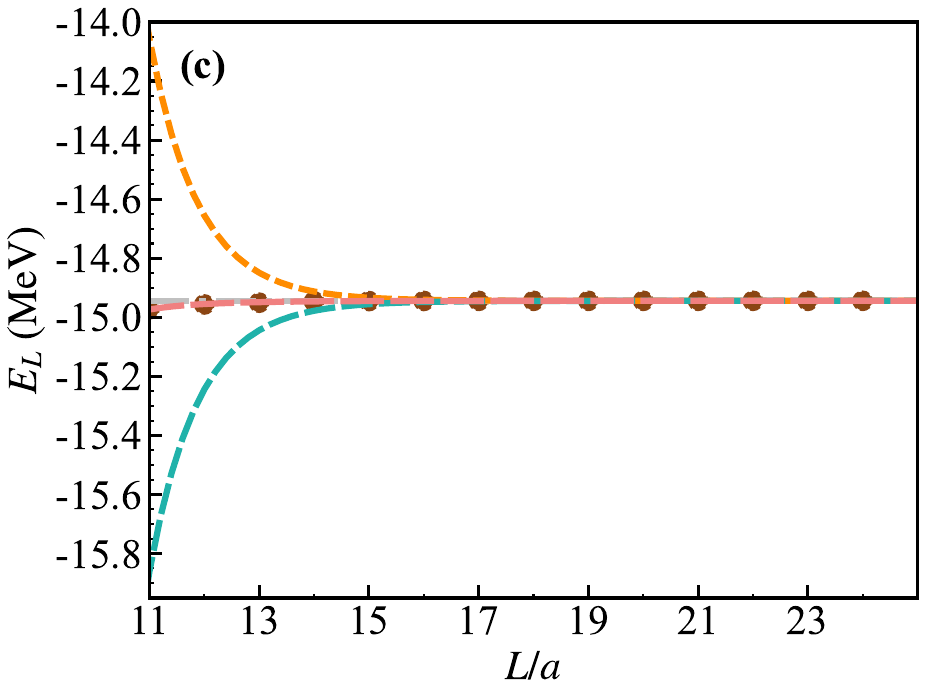}\includegraphics[width=0.4\linewidth]{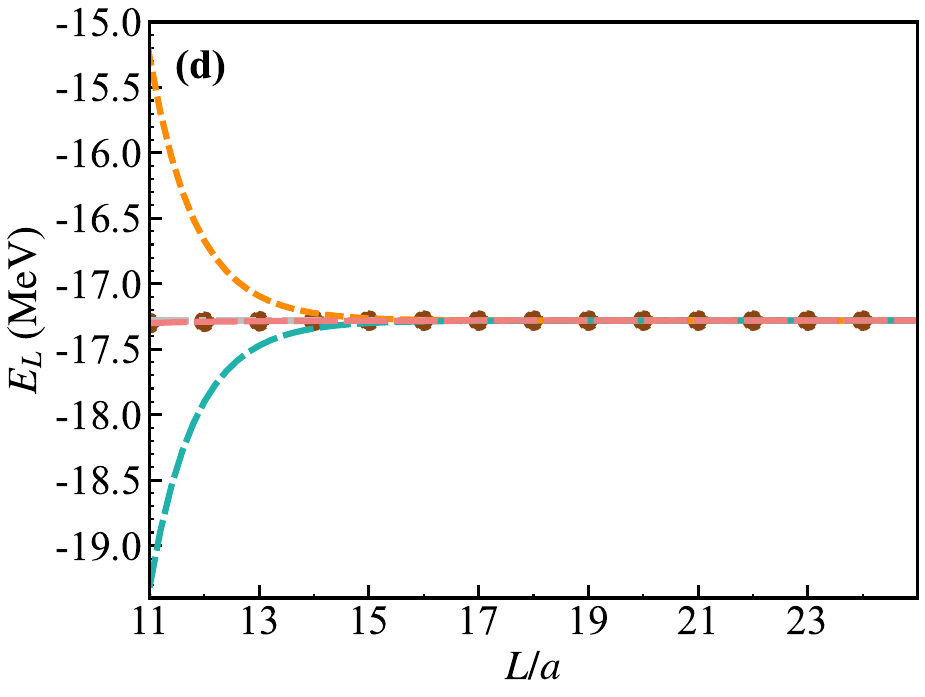}
    \caption{The fitting results for samples with the range of the force range 1 fm. Figures (a), (b), (c), and (d) correspond to samples with potential strength $C_{01}(C_{02}) =$ 1.5, 1.6, 1.7, 1.8 MeV$^{-2}$, respectively. The brown points are the samples generated by LEFT. The pink long-dashed line, orange short-dashed line and seagreen dashed line correspond to Eq.~\eqref{Eq: PySR_3b_Short_Long_Range_Formula}, its short‑range contribution and its long‑range contribution, separately. The gray dashed line is the fitting value of $\hat{\mathcal{A}}_1$, i.e. the energy in infinite volume, in Eq.~\eqref{Eq: PySR_3b_Short_Long_Range_Formula}. The box sizes $L^3 = 10^3,~11^3,\cdots,~24^3$ fm$^3$ are used.}
    \label{Fig:PySR_FVE_1fm_Range}
\end{figure*}
To explore the reliability of our formula, i.e. Eq.~\eqref{Eq: PySR_3b_Short_Long_Range_Formula},  when the samples with a force range tend to be short-range, we specifically generate samples by HLEFT with a force range of about 1 fm and fit them. 
The fitting results are presented in Fig.~\ref{Fig:PySR_FVE_1fm_Range}. 
Here, the pink long-dashed line, orange short-dashed line and seagreen dashed line correspond to Eq.~\eqref{Eq: PySR_3b_Short_Long_Range_Formula}, its short‑range contribution and its long‑range contribution, separately. The gray dashed line is the fitting value of $\hat{\mathcal{A}}_1$ in Eq.~\eqref{Eq: PySR_3b_Short_Long_Range_Formula}. It should be noted that, since the samples satisfies $\kappa_3 \gg \kappa_2$, the short-range contribution in our formula only considers the third term of Eq.~\eqref{Eq: PySR_3b_Short_Long_Range_Formula} for fitting. For all samples of different strength of potential $C_{01}(C_{02})$, Eq.~\eqref{Eq: PySR_3b_Short_Long_Range_Formula} provides well description of these samples. Compared to the fitting results of the samples with force range of 3, 5, 8 fm in Fig.~\ref{Fig:PySR_FVE_Long_range}, the short-range contribution of Eq.~\eqref{Eq: PySR_3b_Short_Long_Range_Formula} significantly increases in 1 fm case.
Due to the short-range contributions, the values of parameters $\hat{\mathcal{A}}_2$ and $\hat{\mathcal{A}}_3$ for the long-range contribution do not follow the rules in Fig.~\ref{Fig:Long_Range_D2E4_D3E5}.
Overall, in order to find a unified formula to describe the finite volume effects of samples under different force ranges, the formula needs to include contributions of both short-range and long-range terms, as shown in Eq.~\eqref{Eq: PySR_3b_Short_Long_Range_Formula}.

\section{Summary and Outlook}\label{sec:Summary}

FVE is a vital step to extract some physical observables for lattice calculations. In two-body system and three-body system, FVE has been rigorously derived in the case of short-range potential. However, challenge exists in long-range potential, especially when the range of the force is comparable to the box size $L$. Several works on two-body system have numerically modified L\"uscher formula to extract the FVE formula but a gap still remains in three-body system. Two-body system has already explored a general formula by SR and the same way are utilized in three-body system.  

In short-range potential, formulae in theory are well rediscovered for two extreme cases, i.e., $\kappa_3\gg\kappa_2$ and $\kappa_3\sim\kappa_2$, and between these two situations, the formula is the liner combination of these two cases, which is greatly influenced by the strength of short-range potential. For the long-range potential, PySR yields the expression given in Eq.~\eqref{Eq: PySR_3b_Long_Range_Formula}. Once short-range potential contributions are included, the formula can be recast into Eq.~\eqref{Eq: PySR_3b_Short_Long_Range_Formula}, which is equivalent to the original PySR solution. In the intermediate force range (approximately 1 fm), Eq.~\eqref{Eq: PySR_3b_Short_Long_Range_Formula} exhibits outstanding performance. Consequently, Eq.~\eqref{Eq: PySR_3b_Short_Long_Range_Formula} can serve as a general formula for describing three-body systems.

Our work makes a notable step forward in the application of data-driven methodologies to explore previously inaccessible regimes in particle physics. By integrating modern learning architectures with physical priors, we demonstrate a viable pathway toward discovering novel signatures that lie beyond the reach of conventional analytical tools.

\section{Acknowledgments}
This work was supported in part  by the  
 National Natural Science Foundation of China  (Grant Nos.~12375073,~12547105, and~12375072); CAS Project for Young Scientists in Basic Research (Grant No. YSBR-117).

\nocite{*}
\bibliography{ref.bib}  

\onecolumngrid
\newpage
\appendix

\clearpage

\section{The Samples from Lattice Effect Field Theory}\label{app:Samples}
For the short-range case, the two-body system samples generated by HLEFT are listed in Tab.\ref{Tab:2b_Short_Range_Samples}. HLEFT is used to generate the $E_L$ in the box size $L^3$ ranging from $5^3$ to $34^3$ fm$^3$. The potential in two-body system is $\delta$-potential, whose strength is controlled by $C_0$ ranging from 2.0 to 2.9 MeV$^{-2}$. Using formula $E_L = E_{\infty} + A_1\frac{e^{-A_2L}}{L}$, two-body system binding energy $E_{2b}$ can be fitted and the binding momentum $\kappa_2$ can be also calculated by $\kappa_{2} = \sqrt{m|E_{2b}|}$. The lattice spacing $a = 1/200$ MeV$^{-1}$ cannot be smaller than the typical hadron size, as hadrons are degrees of freedom in HLEFT. The soft cutoff $\Lambda$ = 350 MeV makes sure that it is much larger than the binding momentum $\kappa$ but smaller than $\pi/a$. In three-body system, only short-range potential in each pair of particles is considered. 
\begin{table}[bp]
    \centering
    \caption{Samples generated by HLEFT for two-body system for short-range case. The mass of each particle is set as 1969 MeV. All the samples are under a certain lattice spacing $a = 1/200$ \text{MeV}$^{-1}\approx0.99$ fm and the soft cutoff $\Lambda$ = 350 MeV. $C_0$ adjusts the strength of $\delta$-potential. Two-body binding energy $E_{2b}$ and binding momentum $\kappa_2$ in infinite volume are also fitted.}
    \begin{tblr}{colspec={p{2cm} X X X X X X X X X X }, row{1}={halign=c}, cells={halign=c, valign=m}}
    \hline
    \hline
    $E_{2b}$(MeV)& -2.244 & -2.879 & -3.560 & -4.280 & -5.035 & -5.820 & -6.631 & -7.466 & -8.322 & -9.197\\
    $\kappa_2$(MeV)&66.471 & 75.291 & 83.724 & 91.800 & 99.569 & 107.049 & 114.265 & 121.246 & 128.008 & 134.569\\
    $C_0(\text{MeV}^{-2})$ &2.0& 2.1 &2.2 & 2.3 & 2.4 &2.5 &2.6 &2.7&2.8&2.9\\
    
    \hline
    $L/a$                  &\SetCell[c=10]{c}$E_L$(MeV)\\
    
5   & -5.785 & -6.265 & -6.767 & -7.293 & -7.841 &-8.412   & -9.006 &-9.622   &-10.260    & -10.920 \\
6   & -4.738 & -5.252 & -5.804 & -6.395 & -7.025 &-7.691   & -8.394 &-9.131   &-9.901    &  -10.701\\
7   & -3.867 & -4.372 & -4.925 & -5.525 & -6.169 &-6.855   & -7.581 &-8.343   &-9.138    &  -9.962\\
8   & -3.216 & -3.705 & -4.247 & -4.840 & -5.481 &-6.166   & -6.889 &-7.648   &-8.438    &  -9.255\\
9   & -2.913 & -3.429 & -4.004 & -4.635 & -5.315 &-6.040   & -6.803 &-7.600   &-8.425    &  -9.275\\
10  & -2.708 & -3.248 & -3.851 & -4.511 & -5.220 &-5.971   & -6.757 &-7.574   &-8.417    &  -9.283\\
11  & -2.545 & -3.102 & -3.723 & -4.399 & -5.122 &-5.883   & -6.677 &-7.500   &-8.347    &  -9.214\\
12  & -2.443 & -3.017 & -3.655 & -4.345 & -5.079 &-5.849   & -6.650 &-7.477   &-8.328    &  -9.199\\
13  & -2.379 & -2.969 & -3.621 & -4.321 & -5.063 &-5.840   & -6.646 &-7.478   &-8.331    &  -9.204\\
14  & -2.334 & -2.936 & -3.596 & -4.304 & -5.051 &-5.830   & -6.639 &-7.472   &-8.326    &  -9.200\\
15  & -2.303 & -2.915 & -3.581 & -4.293 & -5.043 &-5.824   & -6.634 &-7.468   &-8.323    &  -9.197\\
16  & -2.284 & -2.902 & -3.573 & -4.288 & -5.040 &-5.823   & -6.633 &-7.468   &-8.323    &  -9.198\\
17  & -2.271 & -2.894 & -3.568 & -4.285 & -5.038 &-5.821   & -6.632 &-7.467   &-8.323    &  -9.198\\
18  & -2.262 & -2.888 & -3.565 & -4.283 & -5.036 &-5.821   & -6.632 &-7.466   &-8.322    &  -9.197\\
19  & -2.256 & -2.885 & -3.563 & -4.282 & -5.036 &-5.820   & -6.631 &-7.466   &-8.322    &  -9.197\\
20  & -2.252 & -2.883 & -3.562 & -4.281 & -5.035 &-5.820   & -6.631 &-7.466   &-8.322    &  -9.197\\
21  & -2.250 & -2.882 & -3.561 & -4.281 & -5.035 &-5.820   & -6.631 &-7.466   &-8.322    &  -9.197\\
22  & -2.248 & -2.881 & -3.561 & -4.280 & -5.035 &-5.820   & -6.631 &-7.466   &-8.322    &  -9.197\\
23  & -2.247 & -2.880 & -3.560 & -4.280 & -5.035 &-5.820   & -6.631 &-7.466   &-8.322    &  -9.197\\
24  & -2.246 & -2.880 & -3.560 & -4.280 & -5.035 &-5.820   & -6.631 &-7.466   &-8.322    &  -9.197\\
25  & -2.245 & -2.880 & -3.560 & -4.280 & -5.035 &-5.820   & -6.631 &-7.466   &-8.322    &  -9.197\\
26  & -2.245 & -2.879 & -3.559 & -4.280 & -5.035 &-5.820   & -6.631 &-7.466   &-8.322    &  -9.197\\
27  & -2.245 & -2.879 & -3.560 & -4.280 & -5.035 &-5.820   & -6.631 &-7.466   &-8.322    &  -9.197\\
28  & -2.244 & -2.879 & -3.560 & -4.280 & -5.035 &-5.820   & -6.631 &-7.466   &-8.322    &  -9.197\\
29  & -2.244 & -2.879 & -3.560 & -4.280 & -5.035 &-5.820   & -6.631 &-7.466   &-8.322    &  -9.197\\
30  & -2.244 & -2.879 & -3.560 & -4.280 & -5.035 &-5.820   & -6.631 &-7.466   &-8.322    &  -9.197\\
31  & -2.244 & -2.879 & -3.560 & -4.280 & -5.035 &-5.820   & -6.631 &-7.466   &-8.322    &  -9.197\\
32  & -2.244 & -2.879 & -3.560 & -4.280 & -5.035 &-5.820   & -6.631 &-7.466   &-8.322    &  -9.197\\
33  & -2.244 & -2.879 & -3.560 & -4.280 & -5.035 &-5.820   & -6.631 &-7.466   &-8.322    &  -9.197\\
34  & -2.244 & -2.879 & -3.560 & -4.280 & -5.035 &-5.820   & -6.631 &-7.466   &-8.322    &  -9.197\\
    \hline
    \hline
    \end{tblr}
    \label{Tab:2b_Short_Range_Samples}
\end{table}
HLEFT are used to generate $E_L$ in different box sizes $L^3$, and $L/a$ = 7,~8,~9,$\cdots$,24 are utilized as input for PySR. The detail data is listed in Tab.\ref{Tab:3b_Short_Range_Samples}. 
\begin{table}[htbp]
    \centering
    \caption{Samples generated by HLEFT for three-body system for short-range case. The mass of each particle is set as 1969 MeV. All the samples are under a certain lattice spacing $a = 1/200$ \text{MeV}$^{-1}\approx0.99$ fm and the soft cutoff $\Lambda$ = 350 MeV. $C_0$ adjusts the strength of $\delta$-potential. Three-body binding energy $E_{3b}$, fitted by Eq.~\eqref{Eq: PySR_3b_Short_Range_Formula} and binding momentum $\kappa_3$ ($\kappa_3 = \sqrt{m|E_{3b}|}$) in infinite volume are also provided.}
    \begin{tblr}{colspec={p{2cm} X X X X X X X X X X }, row{1}={halign=c}, cells={halign=c, valign=m}}
    \hline
    \hline
    $E_{3b}$(MeV)& -12.003 & -13.790 & -15.614 & -17.469 & -19.352 & -21.259 & -23.187 & -25.134 & -27.097 & -29.076\\
    $\kappa_3$(MeV)&153.733 & 164.780 & 175.340 & 185.463 & 195.203 & 204.595 & 213.671 & 222.461 & 230.985 & 239.271\\
    $C_0(\text{MeV}^{-2})$ &2.0& 2.1 &2.2 & 2.3 & 2.4 &2.5 &2.6 &2.7&2.8&2.9\\
    
    \hline
    $L/a$                  &\SetCell[c=10]{c}$E_L$(MeV)\\
     
7   & -13.688 & -15.336 & -17.048 & -18.814 & -20.624 & -22.472 & -24.352 & -26.260 & -28.193 & -30.146 \\
8   & -12.525 & -14.204 & -15.939 & -17.721 & -19.542 & -21.395 & -23.276 & -25.182 & -27.108 & -29.052 \\
9   & -12.283 & -14.024 & -15.814 & -17.644 & -19.508 & -21.400 & -23.318 & -25.257 & -27.214 & -29.189 \\
10  & -12.142 & -13.907 & -15.715 & -17.559 & -19.434 & -21.335 & -23.259 & -25.203 & -27.164 & -29.141 \\
11  & -12.049 & -13.824 & -15.639 & -17.488 & -19.365 & -21.268 & -23.193 & -25.137 & -27.098 & -29.074 \\
12  & -12.024 & -13.806 & -15.626 & -17.478 & -19.359 & -21.265 & -23.192 & -25.138 & -27.101 & -29.079 \\
13  & -12.014 & -13.799 & -15.621 & -17.475 & -19.357 & -21.264 & -23.191 & -25.138 & -27.102 & -29.080 \\
14  & -12.007 & -13.793 & -15.616 & -17.470 & -19.353 & -21.259 & -23.187 & -25.134 & -27.097 & -29.076 \\
15  & -12.004 & -13.791 & -15.615 & -17.470 & -19.352 & -21.259 & -23.187 & -25.134 & -27.097 & -29.076 \\
16  & -12.004 & -13.791 & -15.614 & -17.469 & -19.352 & -21.259 & -23.187 & -25.134 & -27.097 & -29.076 \\
17  & -12.003 & -13.790 & -15.614 & -17.469 & -19.352 & -21.259 & -23.187 & -25.134 & -27.097 & -29.076 \\
18  & -12.003 & -13.790 & -15.614 & -17.469 & -19.352 & -21.259 & -23.187 & -25.134 & -27.097 & -29.076 \\
19  & -12.003 & -13.790 & -15.614 & -17.469 & -19.352 & -21.259 & -23.187 & -25.134 & -27.097 & -29.076 \\
20  & -12.003 & -13.790 & -15.614 & -17.469 & -19.352 & -21.259 & -23.187 & -25.134 & -27.097 & -29.076 \\
21  & -12.003 & -13.790 & -15.614 & -17.469 & -19.352 & -21.259 & -23.187 & -25.134 & -27.097 & -29.076 \\
22  & -12.003 & -13.790 & -15.614 & -17.469 & -19.352 & -21.259 & -23.187 & -25.134 & -27.097 & -29.076 \\
23  & -12.003 & -13.790 & -15.614 & -17.469 & -19.352 & -21.259 & -23.187 & -25.134 & -27.097 & -29.076 \\
24  & -12.003 & -13.790 & -15.614 & -17.469 & -19.352 & -21.259 & -23.187 & -25.134 & -27.097 & -29.076 \\
    \hline
    \hline
    \end{tblr}
    \label{Tab:3b_Short_Range_Samples}
\end{table}

For the long-range potential case, samples for two-body system and three-body system are also considered and generated by HLEFT. Tab.\ref{Tab:2b_Long_Range_Samples_1and3fm} and Tab.\ref{Tab:2b_Long_Range_Samples_5and8fm} are the samples for two-body system constrained in long-range potential with the force range parameter $\mu = 200,~65,~40,~25$ MeV, which correspond force range about $1,~3,~5,~8$ fm. The lattice spacing $a$ and the soft cutoff $\Lambda$ are the same as those in short-range potential case. $L/a$ ranges from 10 to 49. For $\mu = 200$ MeV case, we adopt $C_{01} = C_{02} = 1.5,1.6,1.7,1.8$ MeV$^{-2}$. For $\mu = 65$ MeV case, we adopt $C_{01} = C_{02} = 1.3,1.6,1.8,2.0$ MeV$^{-2}$. For $\mu = 40$ MeV case, we adopt $C_{01} = C_{02} = 1.2,1.4,1.6,1.8$ MeV$^{-2}$. For $\mu = 25$ MeV case, we adopt $C_{01} = C_{02} = 1.1,1.3,1.4,1.6$ MeV$^{-2}$. Binding energy $E_{2b}$ in infinite volume is fitted by formula, $E_L = E_{\infty} + A_1\frac{e^{-A_2L}}{L^n}$ in our previous work and binding momentum $\kappa_{2}$. 
\begin{table}[htbp]
    \centering
    \caption{Samples generated by HLEFT for two-body system for long-range case for 1 fm and 3 fm force ranges. The mass of each particle is set as 1969 MeV. All the samples are under a certain lattice spacing $a = 1/200$ \text{MeV}$^{-1}\approx0.99$ fm and the soft cutoff $\Lambda$ = 350 MeV. $C_{01}=C_{02}$ adjust the strength of $\delta$-potential and Yukawa potential. Two-body binding energy $E_{2b}$ and binding momentum $\kappa_2$ in infinite volume are also provided.   }
    \begin{tblr}{
    colspec={p{2.2cm} X X X X X X X X},
    width=\linewidth,
    hspan=minimal,   
    row{1}={halign=c}, 
    cells={halign=c, valign=m}
}
    \hline
    \hline
     force range(fm)&1&1&1&1&3&3&3&3\\
   $E_{2b}$(MeV)& -1.541 & -2.206 & -2.949 & -3.757 & -4.266 & -7.795 & -10.457 & -13.295\\
    $\kappa_2$(MeV)&55.084 & 65.906 & 76.201 & 86.009 & 91.650 & 123.888 & 143.492 & 161.796 \\
    $C_{01}/C_{02}(\text{MeV}^{-2})$ &1.5& 1.6 &1.7 & 1.8 & 1.3 &1.6 &1.8 &2.0\\
    
    \hline
    $L/a$                  &\SetCell[c=8]{c}$E_L$(MeV)\\
10&-2.198&-2.728&-3.352&-4.065&-5.509&-8.927&-11.625&-14.539\\
11&-2.006&-2.551&-3.196&-3.931&-5.022&-8.468&-11.155&-14.042\\
12&-1.872&-2.436&-3.103&-3.859&-4.731&-8.207&-10.889&-13.760\\
13&-1.779&-2.361&-3.048&-3.820&-4.557&-8.056&-10.733&-13.595\\
14&-1.711&-2.310&-3.011&-3.794&-4.449&-7.961&-10.634&-13.488\\
15&-1.662&-2.275&-2.988&-3.779&-4.382&-7.902&-10.572&-13.420\\
16&-1.628&-2.253&-2.973&-3.770&-4.341&-7.865&-10.533&-13.378\\
17&-1.603&-2.237&-2.965&-3.765&-4.314&-7.841&-10.507&-13.350\\
18&-1.585&-2.227&-2.959&-3.762&-4.298&-7.826&-10.490&-13.332\\
19&-1.573&-2.220&-2.955&-3.760&-4.287&-7.816&-10.480&-13.320\\
20&-1.564&-2.216&-2.953&-3.759&-4.280&-7.809&-10.472&-13.312\\
21&-1.557&-2.213&-2.951&-3.758&-4.275&-7.805&-10.467&-13.307\\
22&-1.553&-2.211&-2.950&-3.757&-4.272&-7.802&-10.464&-13.303\\
23&-1.550&-2.209&-2.950&-3.757&-4.270&-7.800&-10.462&-13.300\\
24&-1.547&-2.208&-2.949&-3.757&-4.269&-7.798&-10.460&-13.299\\
25&-1.546&-2.208&-2.949&-3.757&-4.268&-7.797&-10.459&-13.298\\
26&-1.544&-2.207&-2.949&-3.757&-4.267&-7.797&-10.459&-13.297\\
27&-1.544&-2.207&-2.949&-3.757&-4.267&-7.796&-10.458&-13.296\\
28&-1.543&-2.207&-2.949&-3.757&-4.267&-7.796&-10.458&-13.296\\
29&-1.542&-2.206&-2.949&-3.757&-4.267&-7.796&-10.458&-13.296\\
30&-1.542&-2.206&-2.948&-3.757&-4.266&-7.795&-10.457&-13.295\\
31&-1.542&-2.206&-2.948&-3.757&-4.266&-7.795&-10.457&-13.295\\
32&-1.542&-2.206&-2.949&-3.757&-4.266&-7.795&-10.457&-13.295\\
33&-1.542&-2.206&-2.949&-3.757&-4.266&-7.795&-10.457&-13.295\\
34&-1.541&-2.206&-2.949&-3.757&-4.266&-7.795&-10.457&-13.295\\
35&-1.541&-2.206&-2.949&-3.757&-4.266&-7.795&-10.457&-13.295\\
36&-1.541&-2.206&-2.949&-3.757&-4.266&-7.795&-10.457&-13.295\\
37&-1.541&-2.206&-2.949&-3.757&-4.266&-7.795&-10.457&-13.295\\
38&-1.541&-2.206&-2.949&-3.757&-4.266&-7.795&-10.457&-13.295\\
39&-1.541&-2.206&-2.949&-3.757&-4.266&-7.795&-10.457&-13.295\\
40&-1.541&-2.206&-2.949&-3.757&-4.266&-7.795&-10.457&-13.295\\
41&-1.541&-2.206&-2.949&-3.757&-4.266&-7.795&-10.457&-13.295\\
42&-1.541&-2.206&-2.949&-3.757&-4.266&-7.795&-10.457&-13.295\\
43&-1.541&-2.206&-2.949&-3.757&-4.266&-7.795&-10.457&-13.295\\
44&-1.541&-2.206&-2.949&-3.757&-4.266&-7.795&-10.457&-13.295\\
45&-1.541&-2.206&-2.949&-3.757&-4.266&-7.795&-10.457&-13.295\\
46&-1.541&-2.206&-2.949&-3.757&-4.266&-7.795&-10.457&-13.295\\
47&-1.541&-2.206&-2.949&-3.757&-4.266&-7.795&-10.457&-13.295\\
48&-1.541&-2.206&-2.949&-3.757&-4.266&-7.795&-10.457&-13.295\\
49&-1.541&-2.206&-2.949&-3.757&-4.266&-7.795&-10.457&-13.295\\
    \hline
    \hline
    \end{tblr}
    \label{Tab:2b_Long_Range_Samples_1and3fm}
\end{table}

\begin{table}[htbp]
    \centering
    \caption{Samples generated by HLEFT for two-body system for long-range case for 5 fm and 8 fm force ranges. Other similar to Table~\ref{Tab:2b_Long_Range_Samples_1and3fm}.}
    \begin{tblr}{
    colspec={p{2.2cm} X X X X X X X X},
    width=\linewidth,
    hspan=minimal,   
    row{1}={halign=c}, 
    cells={halign=c, valign=m}
}
    \hline
    \hline
    force range(fm)&5&5&5&5&8&8&8&8\\
   $E_{2b}$(MeV)& -4.632 & -7.073 & -9.816 & -12.785 & -4.472 & -6.941 & -8.312 & -11.265\\
    $\kappa_2$(MeV)&95.501 & 118.012 & 139.024 & 158.662 & 93.837 &116.905&127.931& 148.932 \\
    $C_{01}/C_{02}(\text{MeV}^{-2})$ &1.2& 1.4 &1.6 & 1.8 & 1.1 &1.3 &1.4 &1.6\\
    
    \hline
    $L/a$                  &\SetCell[c=8]{c}$E_L$(MeV)\\
10&-8.717&-11.472& -14.651 & -18.128 &-16.257&-20.392&-22.650&-27.463\\
11&-7.381&-10.031& -13.083 & -16.406 &-12.880&-16.536&-18.551&-22.854\\
12&-6.517&-9.111 & -12.079 & -15.301 &-10.599&-13.948&-15.800&-19.757\\
13&-5.948&-8.506 & -11.416 & -14.569 &-9.021 &-12.162&-13.898&-17.610\\
14&-5.564&-8.096 & -10.963 & -14.066 &-7.902 &-10.893&-12.546&-16.079\\
15&-5.301&-7.813 & -10.649 & -13.716 &-7.096 &-9.975 &-11.565&-14.967\\
16&-5.118&-7.615 & -10.427 & -13.469 &-6.503 &-9.298 &-10.841&-14.145\\
17&-4.989&-7.473 & -10.269 & -13.292 &-6.062 &-8.791 &-10.298&-13.527\\
18&-4.897&-7.371 & -10.154 & -13.164 &-5.727 &-8.405 &-9.884&-13.057\\
19&-4.830&-7.297 & -10.070 & -13.070 &-5.472 &-8.109 &-9.567&-12.695\\
20&-4.781&-7.242 & -10.008 & -13.000 &-5.273 &-7.879 &-9.320&-12.413\\
21&-4.745&-7.201 & -9.962  & -12.949 &-5.118 &-7.699 &-9.126&-12.193\\
22&-4.718&-7.171 & -9.928  & -12.910 &-4.996 &-7.556 &-8.973&-12.018\\
23&-4.698&-7.148 & -9.902  & -12.881 &-4.899 &-7.442 &-8.851&-11.879\\
24&-4.682&-7.130 & -9.882  & -12.859 &-4.821 &-7.352 &-8.753&-11.767\\
25&-4.671&-7.117 & -9.867  & -12.842 &-4.759 &-7.278 &-8.674&-11.678\\
26&-4.662&-7.107 & -9.855  & -12.829 &-4.708 &-7.219 &-8.610&-11.605\\
27&-4.655&-7.099 & -9.847  & -12.819 &-4.667 &-7.171 &-8.558&-11.546\\
28&-4.650&-7.093 & -9.840  & -12.812 &-4.633 &-7.131 &-8.516&-11.497\\
29&-4.646&-7.089 & -9.835  & -12.806 &-4.606 &-7.099 &-8.481&-11.457\\
30&-4.643&-7.085 & -9.831  & -12.801 &-4.583 &-7.072 &-8.452&-11.425\\
31&-4.641&-7.082 & -9.827  & -12.798 &-4.564 &-7.050 &-8.429&-11.398\\
32&-4.639&-7.080 & -9.825  & -12.795 &-4.548 &-7.031 &-8.409&-11.375\\
33&-4.637&-7.079 & -9.823  & -12.793 &-4.535 &-7.016 &-8.392&-11.356\\
34&-4.636&-7.077 & -9.822  & -12.791 &-4.525 &-7.004 &-8.379&-11.341\\
35&-4.635&-7.076 & -9.820  & -12.790 &-4.516 &-6.993 &-8.367&-11.328\\
36&-4.634&-7.075 & -9.819  & -12.789 &-4.508 &-6.984 &-8.358&-11.317\\
37&-4.634&-7.075 & -9.819  & -12.788 &-4.502 &-6.977 &-8.350&-11.308\\
38&-4.633&-7.074 & -9.818  & -12.788 &-4.496 &-6.970 &-8.343&-11.300\\
39&-4.633&-7.074 & -9.818  & -12.787 &-4.492 &-6.965 &-8.337&-11.294\\
40&-4.633&-7.073 & -9.817  & -12.787 &-4.488 &-6.961 &-8.333&-11.288\\
41&-4.633&-7.073 & -9.817  & -12.786 &-4.485 &-6.957 &-8.329&-11.284\\
42&-4.633&-7.073 & -9.817  & -12.786 &-4.482 &-6.954 &-8.325&-11.280\\
43&-4.633&-7.073 & -9.817  & -12.786 &-4.480 &-6.951 &-8.322&-11.276\\
44&-4.632&-7.073 & -9.817  & -12.786 &-4.478 &-6.949 &-8.320&-11.273\\
45&-4.632&-7.073 & -9.816  & -12.785 &-4.476 &-6.947 &-8.318&-11.271\\
46&-4.632&-7.073 & -9.816  & -12.785 &-4.475 &-6.945 &-8.316&-11.269\\
47&-4.632&-7.073 & -9.816  & -12.785 &-4.474 &-6.944 &-8.314&-11.267\\
48&-4.632&-7.073 & -9.816  & -12.785 &-4.473 &-6.943 &-8.313&-11.266\\
49&-4.632&-7.073 & -9.816  & -12.785 &-4.472 &-6.941 &-8.312&-11.265\\
    \hline
    \hline
    \end{tblr}
    \label{Tab:2b_Long_Range_Samples_5and8fm}
\end{table}
In three-body system, lattice spacing $a$ and the soft cutoff $\Lambda$ are the same as what we have set in two-body system. $L/a$ ranges from 10 to 24. $C_{01} = C_{02}$ adjust the strength of short-range potential and long-range potential separately. The binding energies $E_{3b}$ in force ranges about 1, 3, 5, and 8 fm samples are fitted by Eq.~\eqref{Eq: PySR_3b_Long_Range_Formula} and Eq.~\eqref{Eq: PySR_3b_Short_Long_Range_Formula}. The binding momentum $\kappa_{3} = \sqrt{m|E_{3b}|}$. The samples can be seen in Tab.\ref{Tab:3b_Long_Range_Samples_1and3fm} and Tab.\ref{Tab:3b_Long_Range_Samples_5and8fm}.
\begin{table}[htbp]
    \centering
    \caption{Samples generated by HLEFT for three-body system for long-range case for 1 fm and 3 fm force ranges. Other similar to Table~\ref{Tab:2b_Long_Range_Samples_1and3fm}.}
    \begin{tblr}{
    colspec={p{2.2cm} X X X X X X X X},
    width=1\linewidth,
    hspan=minimal,   
    row{1}={halign=c}, 
    rowsep=1.9pt,
    cells={halign=c, valign=m}
}
    \hline
    \hline
     force range(fm)&1&1&1&1&3&3&3&3\\
   $E_{3b}$(MeV)& -10.458 & -12.666 & -14.944 & -17.281 &-19.860& -30.347 & -37.653 & -45.131\\
    $\kappa_3$(MeV)&143.498 & 157.922 & 171.536 & 184.462 & 197.748 & 244.445 & 272.284 & 298.099 \\
    $C_{01}/C_{02}(\text{MeV}^{-2})$ &1.5& 1.6 &1.7 & 1.8 & 1.3 &1.6 &1.8 &2.0\\
    \hline
    $L/a$           &\SetCell[c=8]{c}$E_L$(MeV)\\
10&-10.622&-12.789&-15.041&-17.361&-22.279&-33.239&-40.882&-48.702\\
11&-10.521&-12.708&-14.973&-17.301&-21.333&-32.119&-39.634&-47.324\\
12&-10.486&-12.684&-14.956&-17.290&-20.783&-31.464&-38.904&-46.517\\
13&-10.471&-12.674&-14.950&-17.286&-20.449&-31.062&-38.454&-46.019\\
14&-10.463&-12.669&-14.946&-17.282&-20.240&-30.809&-38.171&-45.704\\
15&-10.460&-12.667&-14.945&-17.282&-20.109&-30.649&-37.992&-45.507\\
16&-10.459&-12.666&-14.944&-17.282&-20.024&-30.546&-37.877&-45.379\\
17&-10.458&-12.666&-14.944&-17.281&-19.969&-30.480&-37.802&-45.296\\
18&-10.458&-12.666&-14.944&-17.281&-19.933&-30.436&-37.753&-45.241\\
19&-10.458&-12.666&-14.944&-17.281&-19.910&-30.407&-37.721&-45.206\\
20&-10.458&-12.666&-14.944&-17.281&-19.894&-30.387&-37.699&-45.181\\
21&-10.458&-12.666&-14.944&-17.281&-19.883&-30.375&-37.684&-45.165\\
22&-10.458&-12.666&-14.944&-17.281&-19.876&-30.366&-37.675&-45.154\\
23&-10.458&-12.666&-14.944&-17.281&-19.871&-30.360&-37.668&-45.147\\
24&-10.458&-12.666&-14.944&-17.281&-19.868&-30.356&-37.663&-45.142\\
    \hline
    \hline
    \end{tblr}
    \label{Tab:3b_Long_Range_Samples_1and3fm}
\end{table}
\begin{table}[htbp]
    \centering
    \caption{Samples generated by HLEFT for three-body system for long-range case for 5 fm and 8 fm force ranges. Other similar to Table~\ref{Tab:2b_Long_Range_Samples_1and3fm}.}
    \begin{tblr}{
    colspec={p{2.2cm} X X X X X X X X},
    width=\linewidth,
    hspan=minimal,   
    row{1}={halign=c}, 
    rowsep=1.9pt,
    cells={halign=c, valign=m}
}
    \hline
    \hline
     force range(fm)&5&5&5&5&8&8&8&8\\
   $E_{3b}$(MeV)& -20.992 & -28.575 & -36.490 & -44.634 &-20.308&-28.288  & -32.404 &-40.908 \\
    $\kappa_3$(MeV)&203.306 & 237.201 & 268.046 & 296.453 & 199.966 & 236.007 & 252.593 & 283.810 \\
    $C_1/C_2(\text{MeV}^{-2})$ &1.2& 1.4 &1.6 & 1.8 & 1.1 &1.3 &1.4 &1.6\\
    
    \hline
    $L/a$           &\SetCell[c=8]{c}$E_L$(MeV)\\
10&-31.662&-40.928&-50.558&-60.430&-53.610&-67.467&-74.578&-89.047\\
11&-28.250&-36.995&-46.085&-55.411&-44.168&-56.379&-62.655&-75.444\\
12&-26.046&-34.445&-43.183&-52.153&-37.806&-48.895&-54.604&-66.254\\
13&-24.574&-32.739&-41.237&-49.968&-33.384&-43.685&-48.998&-59.854\\
14&-23.567&-31.569&-39.904&-48.470&-30.232&-39.969&-44.998&-55.286\\
15&-22.866&-30.755&-38.975&-47.427&-27.939&-37.265&-42.088&-51.963\\
16&-22.370&-30.178&-38.317&-46.688&-26.242&-35.263&-39.933&-49.501\\
17&-22.014&-29.764&-37.846&-46.158&-24.967&-33.758&-38.313&-47.652\\
18&-21.756&-29.464&-37.503&-45.773&-23.996&-32.612&-37.080&-46.243\\
19&-21.567&-29.245&-37.253&-45.492&-23.248&-31.730&-36.130&-45.159\\
20&-21.428&-29.082&-37.068&-45.284&-22.667&-31.044&-35.392&-44.316\\
21&-21.324&-28.962&-36.930&-45.129&-22.211&-30.506&-34.813&-43.654\\
22&-21.246&-28.871&-36.827&-45.013&-21.850&-30.080&-34.354&-43.131\\
23&-21.188&-28.803&-36.749&-44.926&-21.563&-29.741&-33.989&-42.714\\
24&-21.143&-28.751&-36.690&-44.860&-21.332&-29.470&-33.697&-42.380\\
    \hline
    \hline
    \end{tblr}
    \label{Tab:3b_Long_Range_Samples_5and8fm}
\end{table}

\clearpage
\section{Introduction of PySR Model}\label{app:Introduction}
High-Performance Symbolic Regression in Python and Julia (PySR) is a multi-population evolutionary algorithm with multiple evolutions performed asynchronously. Each population operates the loop of PySR independently and selects the new individual by several mutations and crossovers based on tournament selection. In this work, PySR receives two kinds of inputs $L$ and $E_L$, which lie on both sides of the equation sign, then it will find the formula as a output according to not only the given samples but also some limits set by individuals.

\begin{figure}[hbp]
    \centering
    \includegraphics[width=1.2\linewidth]{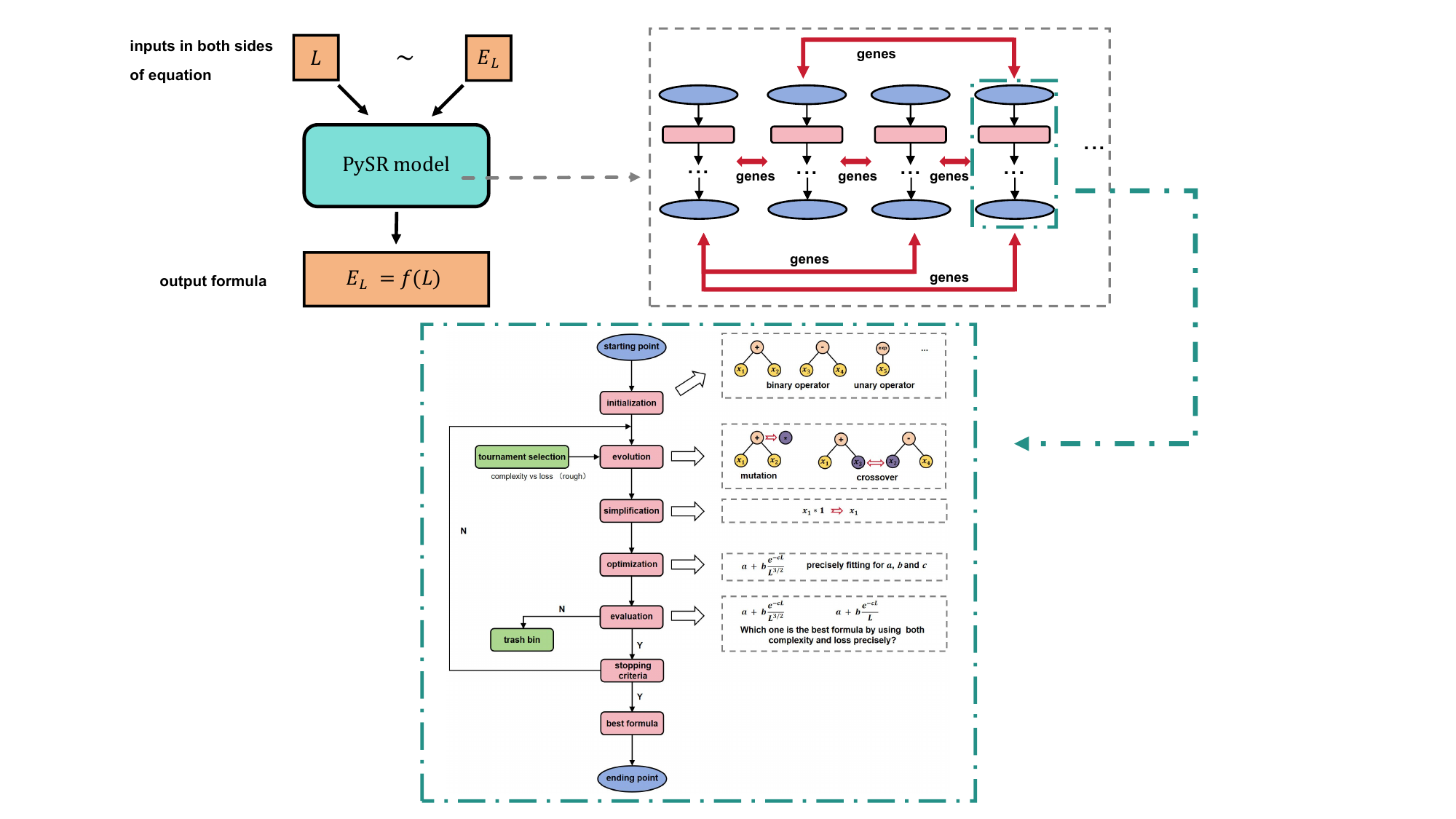}
    \caption{PySR process in detail. The upper left process unenclosed by the dotted line is the most comprehensive process of PySR. The upper right process enclosed by gray dotted line contains several islands in PySR and the genes flow between different islands. The lower process enclosed by Marrs green dot dash line is the evolution process in each island.}
    \label{Fig: PySR_Detail_Process}
\end{figure}

The process of the PySR is modeled after natural evolution. During the initialization stage, PySR produces a series of ``Expression Tree" (some simple formulae) randomly. These trees contain constants, variables and operators (like $+,-,\times,/,\sin,\exp$), which combine together as formulae. Then mutations and crossovers occur in every population, and new formulae are obtained. Before next evolution, PySR will choose the best-performed formula as the next beginning through ``tournament" law, in which the complexity and loss are under rough consideration. For the newly generated formula, some wordy expression like $x\times1$ can be simplified by PySR, which is one of the advantages of PySR compared to traditional SR so that the search can focus on the more effective structure of results. After determining the structure of formula, the unknown parameters in formula will be fitted precisely by optimizer, which produces the most completed formula to evaluation. If the expression of the formula is better than before, it can be accepted and judged whether the process of evolution can be terminated, otherwise the formula is without adoption. Besides, based on traditional SR methods, PySR uses ``Island Model", which means that each island operates a complete set of evolutionary processes mentioned above independently. However, some outstanding genes can be spread between different islands after several generations. Fig.~\ref{Fig: PySR_Detail_Process} expresses these presents in detail.

Some parameters of PySR played a decisive role in the production of the formula. For example, the unary operators must obtains the operator ``$\exp$" term. Different from the two-body case, formula for the short-range case in the three-body system has $L^{-3/2}$ term, therefore, parameter ``extra sympy mappings" needs to construct this term. Besides, the nesting of operators is presented by ``nested constraints". According to the evaluation mechanism of PySR, namely complexity and loss, the complexity of variable is denoted by ``complexity of variables". The complexity of the constants is denoted by ``complexity of constants" and the complexity of the operators is denoted by ``complexity of operators". All the setting of complexity mentioned above can assign specific values. The other evaluation index, loss, is presented by parameter ``elementwise loss". Usually, the loss can be calculated by mean square error (MSE). The selection process of PySR is dominated by parameter ``model selection", which means the chose of formula will consider both complexity and loss. Always, it will be set ``model selection" = best. More parameters are listed in Tab.~\ref{Tab: Parameters with Functions and Values}
\begin{table}[htbp]
\centering
\caption{Some parameters with their functions and values in PySR model}
\begin{tblr}{colspec={p{2.8cm} X p{4cm}}, row{1}={font=\bfseries, halign=c}, cells={halign=c, valign=m}}
\hline\hline
\textbf{Parameter} & \textbf{Function} & \textbf{Value} \\
\hline
populations & determine the number of populations in each generation during evolution & 200 \\
population\_size & number of the candidate in each population & 100\\
ncycles\_per\_iteration & number of total mutations to run, per 10 samples of population, per iteration & 800\\
niterations & number of iterations of the algorithm to run & 200\\
maxsize & upper complexity limit of final result & 30\\
extra\_sympy\_mappings & custom operator not provided & ``func", ``inv"\footnotemark[1] \\
binary\_operators & operators which take two scalar as inputs & ``+", ``$\times$", ``/"\\
unary\_operators & operators which only take a single scalar as input & ``neg", ``exp", ``sqrt",
``cube", ``func"\\
nested\_constraints & specify how many times a combination of operators can be nested & ``exp":\{``exp:0",``func":0.$\cdots$\}\footnotemark[2], ``sqrt":\{``sqrt":0,``+":0,``func":0\}, ``cube":\{``cube":0,``func":0\}, ``/":\{``+":0\}, ``func":\{``func":0,``+":0\}\\
complexity\_of\_variables & global complexity of variables & 1\\
complexity\_of\_constants & complexity of constants & 3\\
complexity\_of\_operators & complexity of different operators &``exp":0, ``func":0, ``cube":0, ``sqrt":0, ``$\times$":1.5\\
optimizer\_iterations & number of iterations that the constants optimizer can take & 15\\
perturbation\_factor & constants are perturbed by a max factor of (perturbation factor$\times$T + 1), either multiplied by this or divided by this & 0.5\\
parsimony & multiplicative factor for how much to punish complexity & 0.005\\
adaptive\_parsimony\_scaling & weigh of simple formula & 15.0\\
model\_selection & select a final expression from the list of best expression at each complexity & ``best"\\
elementwise\_loss & elementwise loss function & ``L2DistLoss"\\

\hline\hline
\end{tblr}

\footnotetext[1]{The complete form of definition: ``func": sqrt(abs(x))$^3$, `"inv": 1/x}
\footnotetext[2]{The complete form of definition: ``exp":\{``exp:0", ``func":0, ``sqrt":0, ``cube":0, ``$\times$":1, ``func":0\}}
\label{Tab: Parameters with Functions and Values}
\end{table}

\clearpage
\section{PySR Results}\label{app:PySR}
\begin{figure}[hbp]
    \centering
    \includegraphics[width=0.5\linewidth]{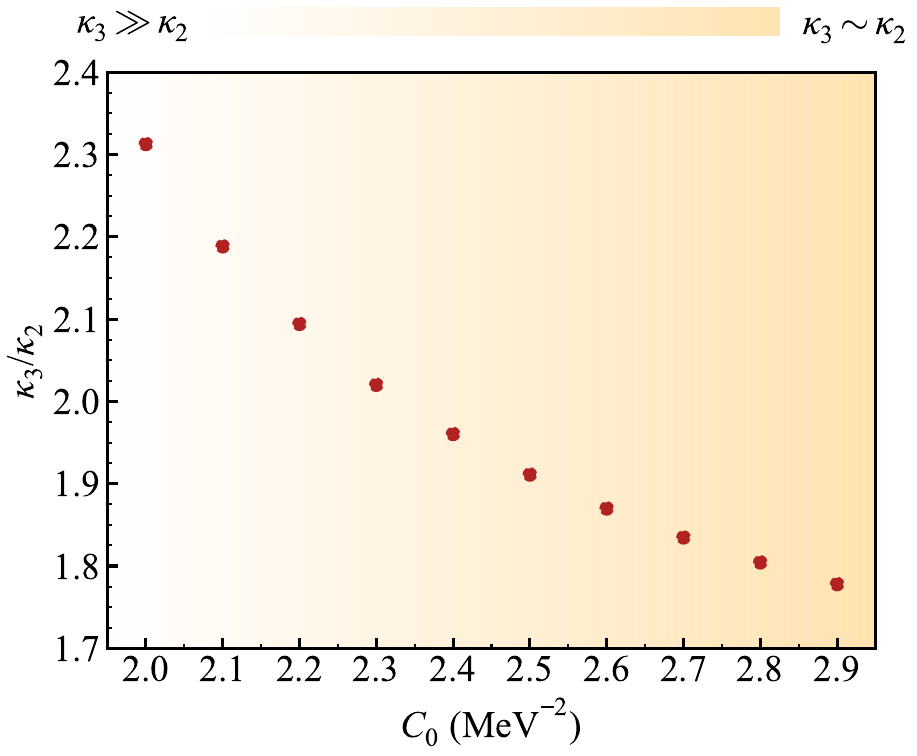}
    \caption{Relationship between $C_0$ and $\kappa_3/\kappa_2$. The brown points represent the ratio of binding momentum $\kappa_3/\kappa_2$ in different $C_0$ ranging from 2.0 to 2.9 $\mathrm{MeV}^{-2}$. The background color approaches orange means $\kappa_3\sim \kappa_2$.}
    \label{Fig: C0_and_k3k2}
\end{figure}

PySR model presents us not only formulae but also other physical results. In short-range case, we have already get Eq.~\eqref{Eq: PySR_3b_Short_Range_Formula}. The $n$ of third term, the importance ratio of $|A_4/A_2|$ and the power of exponential $A_3$ have correlation with the strength of $\delta$-potential $C_0$. In addition, the correlation between momentum ratio of $\kappa_3/\kappa_2$ and $C_0$ has also been calculated and can be read in Tab.\ref{Tab:C0_n_k3/k2_C4/C2_C3} and Fig.~\ref{Fig: C0_and_k3k2}.
\begin{table}[htbp]
    \centering
    \caption{Correlation between $C_0$ and $\kappa_3/\kappa_2$, $n$, $|A_4/A_2|$ and $A_3$ for short-range potential cases.}
    \begin{tblr}{
    colspec={p{1.5cm} X X X X X X X X X X},
    width=\linewidth,
    hspan=minimal,   
    row{1}={halign=c}, 
    cells={halign=c, valign=m}
}
    \hline
    \hline
    $C_0$(MeV$^{-2}$)&2.0&2.1&2.2&2.3&2.4&2.5&2.6&2.7&2.8&2.9\\
   $\kappa_3/\kappa_2$&2.31 & 2.19 & 2.09 & 2.02 & 1.96 & 1.91 & 1.87 & 1.84 & 1.80 & 1.78\\
   $n$&7.08(43) & 6.10(36) & 3.44(17) & 2.27(6) & 1.81(3) & 1.76(2) & 1.36(1) & 1.16(3) & 1.13(3) & 1.03(4)\\
   $|A_4/A_2|$&0.00(0) & 0.00(0) & 0.00(0) & 0.05(2) & 0.31(4) & 0.38(4) & 1.71(10) & 3.62(52) & 4.03(60) & 5.85(110)\\
   $A_3(\mathrm{MeV})$ &148.67 (3251) & 160.40 (3221) & 183.60 (2806) & 201.04 (2733) & 214.28 (2793) & 189.45 (1368) & 230.78 (3011) & 233.85 (3110) & 240.28 (3116) & 243.36 (3219)\\
    \hline
    \hline
    \end{tblr}
    \label{Tab:C0_n_k3/k2_C4/C2_C3}
\end{table}

In long-range force range cases, PySR model presents Eq.~\eqref{Eq: PySR_3b_Long_Range_Formula}
Then short-range potential terms are added and the formula is Eq.~\eqref{Eq: PySR_3b_Short_Long_Range_Formula}.
In long-range cases, different force ranges and their parameters of these two formulae are listed in Tab.~\ref{Tab:C01=C02_D2D4_E2E3E4E5}. 
\begin{table}[htbp]
    \centering
    \caption{Results for different force ranges and force strength of $C_{01}(C_{02})$. Parameters $\mathcal{A}_2$ and $\mathcal{A}_3$ in Eq.~\eqref{Eq: PySR_3b_Long_Range_Formula}. Parameters $\hat{\mathcal{A}}_2,\hat{\mathcal{A}}_3,\hat{\mathcal{A}}_4$ and $\hat{\mathcal{A}}_5$ in Eq.\eqref{Eq: PySR_3b_Short_Long_Range_Formula}.}
    \begin{tblr}{
    colspec={p{2.8cm} X X X X },
    width=\linewidth,
    hspan=minimal,   
    row{1}={halign=c}, 
    cells={halign=c, valign=m}
}
    \hline
    \hline
    force range(fm)&\SetCell[c=4]{c} 3\\
    $C_{01}/C_{02}$(MeV$^{-2}$)&1.3&1.6&1.8&2.0\\
    \hline
    $\mathcal{A}_2(\mathrm{MeV}^{-5/4})$&-0.04&-0.05&-0.06&-0.06\\
    $\mathcal{A}_3(\mathrm{MeV})$&54.37&53.79&53.67&53.61\\
    $\hat{\mathcal{A}}_2(\mathrm{MeV}^{-5/4})$&-0.04&-0.05&-0.06&-0.06\\
    $\hat{\mathcal{A}}_3(\mathrm{MeV})$&54.35&54.13&54.05&54.0\\
    $\hat{\mathcal{A}}_4(\mathrm{MeV}^{-1/2})$&-267.70&-4743.07&-26640.88&-132948.81\\
    $\hat{\mathcal{A}}_5(\mathrm{MeV}^0)$&320.08&3285.39&12328.08&41419.10\\
    \hline
    force range(fm)&\SetCell[c=4]{c} 5\\
    $C_{01}/C_{02}$(MeV$^{-2})$&1.2&1.4&1.6&1.8\\
    \hline
    $\mathcal{A}_2(\mathrm{MeV}^{-5/4})$&-0.067&-0.075&-0.086&-0.096\\
    $\mathcal{A}_3(\mathrm{MeV})$&33.54&32.87&32.81&32.78\\
    $\hat{\mathcal{A}}_2(\mathrm{MeV}^{-5/4})$&-0.063&-0.073&-0.083&-0.094\\
    $\hat{\mathcal{A}}_3(\mathrm{MeV})$&32.47&32.42&32.43&32.43\\
    $\hat{\mathcal{A}}_4(\mathrm{MeV}^{-1/2})$&-20.69&917.11& 8510.04&60875.87\\
    $\hat{\mathcal{A}}_5(\mathrm{MeV}^0)$&-243.97&1762.74&-8170.65&-33750.36\\
    \hline
    force range(fm)&\SetCell[c=4]{c} 8\\
    $C_{01}/C_{02}$(MeV$^{-2})$&1.1&1.3&1.4&1.6\\
    \hline
    $\mathcal{A}_2(\mathrm{MeV}^{-5/4})$&-0.12&-0.15&-0.15&-0.17\\
    $\mathcal{A}_3(\mathrm{MeV})$&22.39&23.09&21.77&22.26\\
    $\hat{\mathcal{A}}_2(\mathrm{MeV}^{-5/4})$&-0.11&-0.13&-0.14&-0.16\\
    $\hat{\mathcal{A}}_3(\mathrm{MeV})$&21.05&21.43&21.18&21.23\\
    $\hat{\mathcal{A}}_4(\mathrm{MeV}^{-1/2})$&3333.15&20624.56&99740.08&740155.35\\
    $\hat{\mathcal{A}}_5(\mathrm{MeV}^0)$&-5461.59& -22159.10&-77866.93&-365810.18\\
    \hline
    \hline
    \end{tblr}
    \label{Tab:C01=C02_D2D4_E2E3E4E5}
\end{table}
Note that parameters $\hat{\mathcal{A}}_4$ and $\hat{\mathcal{A}}_5$ in Eq.~\eqref{Eq: PySR_3b_Short_Long_Range_Formula} are much larger than $\hat{\mathcal{A}}_2$ while the values in the exponent $\frac{2}{\sqrt{3}}\kappa_3$ and $\frac{2}{\sqrt{3}}\sqrt{\kappa_3^2-\kappa_2^2}$ are sufficiently large to completely offset the influence of $\hat{\mathcal{A}}_4$ and $\hat{\mathcal{A}}_5$.    

For 1 fm case, Eq.~\eqref{Eq: PySR_3b_Short_Long_Range_Formula} is utilized for fitting. The fitting result can be read in Tab.~\ref{Tab:C01=C02_D2D3_E2E4E5}.
\begin{table}[htbp]
    \centering
    \caption{Results for 1 fm case of $C_{01}(C_{02})$. Parameters $\hat{\mathcal{A}}_2,\hat{\mathcal{A}}_3$ and $\hat{\mathcal{A}}_4$ in Eq.\eqref{Eq: PySR_3b_Short_Long_Range_Formula}.}
    \begin{tblr}{
    colspec={p{2.8cm} X X X X },
    width=\linewidth,
    hspan=minimal,   
    row{1}={halign=c}, 
    cells={halign=c, valign=m}
}
    \hline
    \hline
    force range(fm)&\SetCell[c=4]{c} 1\\
    $C_{01}/C_{02}$(MeV$^{-2})$&1.5&1.6&1.7&1.8\\
    \hline
    $\hat{\mathcal{A}}_2(\mathrm{MeV}^{-5/4})$&-2.49&-6.15&-33.65&-169.97\\
    $\hat{\mathcal{A}}_3(\mathrm{MeV})$&149.64&166.74&184.19&199.08\\
    $\hat{\mathcal{A}}_4(\mathrm{MeV}^{-1/2})$&45.29&114.47& 615.57&3186.15\\
    \hline
    \hline
    \end{tblr}
    \label{Tab:C01=C02_D2D3_E2E4E5}
\end{table}

\balance
\end{document}